\documentclass[preprint,12pt]{elsarticle}

\usepackage{listings}
\usepackage{graphicx}
\usepackage{booktabs} 
\usepackage{array,multirow}
\usepackage[ruled,noend]{algorithm2e} 
\usepackage{url} 
\usepackage{natbib}
\usepackage{amssymb}
\usepackage{lscape}
\usepackage{mathrsfs}
\usepackage{subfig}
\usepackage{amsmath}
\usepackage{tikz}
\usepackage{float}
\usepackage{rotating}

\usepackage{hyperref}
\usepackage{tikz}
\usetikzlibrary{arrows,shapes,positioning,shadows,trees}

\tikzset{
  basic/.style  = {draw, text width=4cm, drop shadow, font=\sffamily, rectangle},
  root/.style   = {basic, rounded corners=2pt, thin, align=center,fill=green!60},
  level 2/.style = {basic, rounded corners=6pt, thin,align=center, fill=green!30,text width=8em},
  level 3/.style = {basic, thin, align=left, fill=pink!60, text width=6.5em},
  level 3a/.style = {basic, thin, align=left, fill=pink!30, text width=6.5em},
  level 4/.style = {basic, thin, align=left, fill=gray!40, text width=6.5em},
  level 5/.style = {basic, thin, align=center, fill=gray!5,text width=6.5em}
}

\makeatletter
\providecommand{\doi}[1]{%
	\begingroup
	\let\bibinfo\@secondoftwo
	\urlstyle{rm}%
	\href{http://dx.doi.org/#1}{%
		doi:\discretionary{}{}{}%
		\nolinkurl{#1}%
	}%
	\endgroup
}
\makeatother

\usepackage{amssymb}

\journal{Swarm and Evolutionary Computation}

\begin{document}

\begin{frontmatter}



\title{Genetic-based optimization in fog computing: current trends and research opportunities}


\author{Carlos Guerrero\corref{mycorrespondingauthor}}
\ead{carlos.guerrero@uib.es}
\cortext[mycorrespondingauthor]{Corresponding author}

\author{Isaac Lera\corref{}}
\ead{isaac.lera@uib.es}

\author{Carlos Juiz\corref{}}
\ead{cjuiz@uib.es}

\address{Crta. Valldemossa km 7.5, Palma, E07121, SPAIN}

\address[mymainaddress]{Computer Science Department, University of Balearic Islands}

\begin{abstract}

Fog computing is a new computational paradigm that emerged from the need to reduce network usage and latency in the Internet of Things (IoT). Fog can be considered as a continuum between the cloud layer and IoT users that allows the execution of applications or storage/processing of data in network infrastructure devices. The heterogeneity and wider distribution of fog devices are the key differences between cloud and fog infrastructure. Genetic-based optimization is commonly used in distributed systems; however, the differentiating features of fog computing require new designs, studies, and experimentation. The growing research in the field of genetic-based fog resource optimization and the lack of previous analysis in this field have encouraged us to present a comprehensive, exhaustive, and systematic review of the most recent research works. Resource optimization techniques in fog were examined and analyzed, with special emphasis on genetic-based solutions and their characteristics and design alternatives. We defined a taxonomy of the optimization scope in fog infrastructures and used this optimization taxonomy to classify the 70 papers in this survey. Subsequently, the papers were assessed in terms of genetic optimization design. Finally, the benefits and limitations of each surveyed work are outlined in this paper. Based on these previous analyses of the relevant literature, future research directions were identified. We concluded that more research efforts are needed to address the current challenges in data management, workflow scheduling, and service placement. Additionally, there is still room for improved designs and deployments of parallel and hybrid genetic algorithms that leverage, and adapt to, the heterogeneity and distributed features of fog domains.

\end{abstract}


\end{frontmatter}



\section{Introduction}

%
%
%
%
%

Internet of Things (IoT) technologies emerge thanks to the increase in the number of user devices as well as an increase in distributed processing capabilities. Additionally, the IoT technologies have evolved, resulting in the emergence of new computing paradigms such as fog computing. Fog computing is based on the idea of bringing computing and storage resources closer to the point where they are requested or generated, i.e., closer to users. To achieve this, a computing continuum is defined between the cloud and users, because computing resources are incorporated into the intermediate elements of the infrastructure (fog devices). Thus, data can be stored in nearby devices, and services can be executed in nodes closer to the user, reducing latency and network load, and fog devices provide data in motion capabilities~\citep{10.1145/3301443}.

In fog computing, resource management becomes more complex with an increase in the number of nodes involved, their heterogeneity, geographic dispersion, and the asymmetry in the interconnected networks~\citep{Naeem2019}. An adequate resource management strategy should seek the optimization of certain non-functional aspects of the architecture. However, we are faced with an NP-complete and multi-objective problem~\citep{7919155} that must be solved using an optimization method. The studies included in this systematic analysis showed that the use of genetic algorithms (GAs) is a common and adequate solution to implement these resource management processes. Nevertheless, there are many research avenues open for improving GA-based solutions.

We include an comprehensive analysis of the state of the art in this field: the use of GAs to optimize the resource management process of fog architectures. This study offers a twofold analysis of the field of genetic optimization of fog resources: the optimization scope and the design of the GA. 
Two taxonomies, one for each case, were created by analyzing the papers, and they are organized in terms of these taxonomies. 
The main research challenges and direction for future research are identified and explained by analyzing the papers in terms of each element of the taxonomy.

\subsection{Motivation behind the research}

Fog computing has emerged as a promising technology that overcomes the limitations of cloud computing for real-time and data-massive applications. These applications require short response times and generate a large quantity of data that must be transmitted through the infrastructure. The processing and storage capacities of fog devices reduce the response time and network usage; however, new challenges emerge from the heterogeneity and geo-distribution of these devices and their interconnecting networks. Traditional solutions in the field of cloud computing must be revisited or adapted for suitable applications in fog infrastructures. A large number of research papers have already addressed these new challenges and research problems by studying various optimization algorithms and methods. This volume of research justifies the analysis of their contributions to this literature review.

GA is one of the most popular optimization techniques. It is a meta-heuristic inspired by natural selection. The optimization of a GA is based on successive generations that combine (crossover operator) the best solutions (selection operator and fitness) to create new solutions. Random changes (mutation operators) are also considered to avoid local optimization. GA is classified into the group of evolutionary algorithms (EA). The optimization algorithms included in EA differ in genetic representation and other implementation details, as well as the nature of the applied problem. We have limited the survey to papers that used genetic approaches because extending the analysis to all types of EAs would significantly increase the number of papers presented. 

The systematic analysis in this study was motivated by the huge number of contributions that applied GA to optimize fog infrastructure, and the lack of previous analyses of using GA in infrastructure optimization. 

Previous related surveys analyzed the state of the art by exploring the challenges addressed by fog computing~\cite{Naeem2019, 10.1155/2020/8849181}, enabling architectures/technologies~\cite{CAIZA2020e03706, fi12110190, BITTENCOURT2018134, 9194714, 8100873, 10.1145/3326066, ALLI2020100177, cardellini2019self}, application domains~\cite{10.1145/3301443, fi12110190, BITTENCOURT2018134, ALLI2020100177, BILAL201894, 8444370}, security~\cite{CAIZA2020e03706, 9194714}, algorithms/tools~\cite{8100873, 10.1145/3326066, ALLI2020100177}, and/or the network aspects~\cite{8542693}. In addition, there is a set of specific surveys that analyze the use of fog computing in specific domains such as smart homes~\cite{RAHIMI2020102531} or smart cities~\cite{songhorabadi2020fog}. 

Because our survey is focused on optimization techniques for fog computing, we analyzed previous related surveys focused on general optimization scopes. Table~\ref{tab:summaryofsurveys} shows a summary of all the surveys we analyzed, considering the infrastructure domain, optimization scope, and optimization technique. Note that our study is the first to survey genetic-based optimizations in the fog infrastructure domain. 

The most similar survey in Table~\ref{tab:summaryofsurveys} is the \citeauthor{OGUNDOYIN2021100937}'s~\cite{OGUNDOYIN2021100937}. If we compare it with our study, by considering the list of analyzed papers, our study includes 70 papers and \citeauthor{OGUNDOYIN2021100937}'s study includes 138 papers, but only 16 papers overlap between them~\cite{guerrero2019evaluation,8701449,DEMAIO2020171,AbbasiPK20,10.1016/j.future.2019.09.039,10.1145/3287921.3287984,app9091730,LiZSS20,9140118,10.1002/dac.4652,10.1007/s11277-017-5200-5,REDDY2020102428,8947936,8676263,8929234,a12100201}.

\begin{table}
\caption{Analysis of related surveys.}
\label{tab:summaryofsurveys}
\centering
\tabcolsep=0.07cm
\footnotesize
\begin{tabular}{p{3cm}|l|l|l}\toprule

		&	Infrastructure	&	Optimization	&	Optimization	\\
		&	domain	&	scope	&	technique	\\ \midrule
								
\citeauthor{kashani2020load}~\cite{kashani2020load}&	fog	&	load balancing	&	no restrictions	\\
\citeauthor{kaur2021systematic}~\cite{kaur2021systematic}&	fog	&	load balancing	&	no restrictions	\\
\citeauthor{9012677}~\cite{9012677}&	container-based fog	&	orchestration	&	no restrictions	\\
\citeauthor{10.1155/2020/8964165}~\cite{10.1155/2020/8964165}&	cloud, fog, edge	&	quality of service	&	no restrictions	\\
\citeauthor{10.1007/s10115-016-0951-y}~\cite{10.1007/s10115-016-0951-y}&	cloud	&	resource allocation	&	no restrictions	\\
\citeauthor{9144705}~\cite{9144705}&	fog	&	resource allocation	&	no restrictions	\\
\citeauthor{tocze2018}~\cite{tocze2018}&	edge	&	resource management	&	no restrictions	\\
\citeauthor{OGUNDOYIN2021100937}~\cite{OGUNDOYIN2021100937}&	fog	&	resource management	&	no restrictions	\\
\citeauthor{ghobaei2020resource}~\cite{ghobaei2020resource}&	fog	&	resource management	&	no restrictions	\\
\citeauthor{MANVI2014424}~\cite{MANVI2014424}&	iaas cloud	&	resource management	&	no restrictions	\\
\citeauthor{xu2020dynamic}~\cite{xu2020dynamic}&	cloud, fog, edge	&	resource provisioning	&	no restrictions	\\
\citeauthor{10.1145/3326540}~\cite{10.1145/3326540}&	fog, edge	&	service migration	&	no restrictions	\\
\citeauthor{9197885}~\cite{9197885}&	fog	&	service placement	&	no restrictions	\\
\citeauthor{10.1145/3403955}~\cite{10.1145/3403955}&	fog	&	service placement	&	no restrictions	\\
\citeauthor{brogi2020place}~\cite{brogi2020place}&	fog	&	service placement	&	no restrictions	\\
\citeauthor{10.1145/3391196}~\cite{10.1145/3391196}&	fog, edge	&	service placement	&	no restrictions	\\
\citeauthor{HOUSSEIN2021100841}~\cite{HOUSSEIN2021100841}&	cloud	&	task scheduling	&	meta-heuristics	\\
\citeauthor{barros2020scheduling}~\cite{barros2020scheduling}&	cloud, fog	&	task scheduling	&	no restrictions	\\
\citeauthor{abdulredha2020heuristic}~\cite{abdulredha2020heuristic}&	cloud, fog	&	task scheduling	&	meta-heuristics	\\
\citeauthor{varshney2020characterizing}~\cite{varshney2020characterizing}&	cloud, fog, edge	&	task scheduling	&	no restrictions	\\
\citeauthor{8748596}~\cite{8748596}&	fog	&	task scheduling	&	no restrictions	\\
\citeauthor{8862651}~\cite{8862651}&	fog	&	task scheduling	&	no restrictions	\\
\citeauthor{10.1002/dac.4583}~\cite{10.1002/dac.4583}&	fog	&	task scheduling	&	no restrictions	\\
\citeauthor{10.1002/ett.3792}~\cite{10.1002/ett.3792}&	fog	&	task scheduling	&	no restrictions	\\
\citeauthor{10.1145/3325097}~\cite{10.1145/3325097}&	cloud, fog	&	workflow scheduling	&	no restrictions	\\
&&&\\
This work & fog & resource management & GA \\

\bottomrule
\end{tabular}
\end{table}

\subsection{Contributions}

The work presented in this paper is justified by the ongoing research in the field of genetic-based fog resource optimization and the lack of previous analysis in this field. To catch up with this growth, we present a comprehensive, exhaustive, and systematic review of up-to-date research works. The contributions of this study can be summarized as follows.

\begin{enumerate}
    \item Resource optimization techniques in fog computing are examined and analyzed, with a special emphasis on GA, its characteristics, and design alternatives.
    \item A taxonomy for resource optimization in fog is presented, and studies are analyzed in terms of this taxonomy.
    \item The genetic optimization design of existing approaches are assessed. 
    \item Benefits and limitations/weaknesses of each surveyed work are outlined.
    \item The current research challenges and future research directions are presented.
\end{enumerate}

We believe that these contributions will provide researchers insight into the present research efforts and opportunities to explore the identified research challenges.

\subsection{Organization of the paper}

The remainder of this paper is organized as follows. Section~\ref{researchdomain} provides a brief overview of the domains covered in this systematic review. Section~\ref{sec_methodology} describes the methodology used. The analyses of the papers in terms of the optimization scope is presented in Section~\ref{sec:ambitodelsurvey}. Section~\ref{sec:surveyAG} includes an analysis from the perspective of GA design. Finally, future research directions obtained from the analysis are presented in Section~\ref{sect:researchchallenges}.

\section{Brief overview of the research domain}
\label{researchdomain}

This section provides a brief overview of the context of this study. Among the three main domains, resource optimization, fog computing, and GA, we consider that it is worthless to include a brief introduction to GA because it is a well-known domain in the field of evolutionary algorithms and they are deeply explained in related literature~\cite{goldberg2006genetic}. The other two domains are briefly explained in the following subsections.

\subsection{Resource optimization}

\label{optimizacionrecursos}


Resource optimization refers to the efficient management of available resources with finite capacity. The optimization task does a decision-making process in which it will assign the requests to the resources trying to optimize predetermined criteria. Often, these criteria are opposed, causing that the improvement of one will mean that the others will worsen~\citep{KONAK2006992}.


These types of problems are addressed through multi-objective optimization. The solutions will satisfy the objectives in a balanced way, without being dominated by other solutions. The improvement of a given objective in a multi-objective proposal is usually lower than the optimization of the single objective.



A multi-criteria optimization is a $n$-dimensional decision vector of variables $\vec{x}=\{x_1,\ldots,x_n\}$ that belong to the  $n$-dimensional space of solutions  $X^n \subset \mathbb{R} ^{n}$, $\vec{x} \in X^n$~\citep{KONAK2006992, deb2001multi, deb2014multi} 


Not all solutions may be feasible. These can be, first of all, limited by upper and lower limits, such that $x_i^{lower} \leq x_i \leq x_i^{upper}, i=1,\dotsc,n$. In addition, the solutions can be conditioned to meet a set of restrictions, these being both inequality constraints $g_k(\vec{x})\geq 0, k=1,\ldots,o$ and equality constraints, $h_l(\vec{x}) = 0, l=1,\ldots,p$.



The set of solutions that satisfy these constraints and limits establish the space of viable solutions, $S^n \subset X^n$.


Each of the $\vec{x}$ solutions of $S^n$ are evaluated through an objective function for each of the $m$ optimization criteria, $f_j(\vec{x}), j=1,\dotsc,m$, giving place a vector $\vec{z}$ belonging to the objective space $m$-dimensional $Z^m \subset \mathbb{R} ^{m}$, $\vec{z} \in Z^m$, such that  $\vec{z}=\{z_1,\ldots,z_m\}$ where $z_j = f_j(\vec{x}), l=1,\dotsc,m$. In this way, each solution $\vec{x}$ in the solution space is related to another point $\vec{z}$ in the objective space.

Therefore, a multi-objective optimization problem is defined as:

\begin{eqnarray*}
	Minimize/Maximize & f_j(\vec{x}) & j=1,\ldots,m; \\
	subject\ to  &   x_i^{lower} \leq x_i \leq x_i^{upper} & i=1,\dotsc,n; \\
	& g_k(\vec{x})\geq 0 & k=1,\ldots,o; \\
	&  h_l(\vec{x})= 0 & l=1,\ldots,p.
\end{eqnarray*}





To compare the goodness of the solutions, it is necessary to establish an ordering relationship on the points of the objective space. In the case of optimization problems with a single objective, a total order relationship determined by the scalar values of the \emph{fitness} function is established. In multi-objective optimization problems, the individual \textit {fitness} functions are defined for each objective to be minimized, as already commented above, thus having a vector of values of a size equal to the number of objectives to optimize, $\vec{z}=\{z_1,\ldots,z_m\}$.


If we want to establish a total ordering relationship on $\vec{z}$ values, one of the simplest ways is by applying a transformation function with weights ($\vec{\omega}=\{\omega_1,\ldots,\omega_m\}$)  to each of the values of the vector, so that an easily comparable scalar value $e$ is obtained~\citep{marler2004survey}:

\[  e = \omega_1 \cdot z_1 + \ldots + \omega_{m} \cdot z_{m} \]


Alternatively, the dominance concept can be used to order the solutions. A solution $\vec{x}_a$ dominates another solution $\vec{x}_b$ ($\vec{x}_a \prec =\ \vec{x}_b$) if it is better or equal for all the objectives to be optimized, that is for all the values of the vector $\vec{z}_a$ being better in at least one of them. Thus, we can consider the set of non-dominated solutions as the solutions that are not dominated by any other solution in the solution space, i.e., the optimized (or \textit{best}) solutions. This set of solutions in a domain $X^n$ is known as the Pareto set and its image in the objective space $Z^m$ is often called the Pareto front~\cite{coello2007evolutionary}.

Once the Pareto set is determined, the same process can be repeated iteratively, by removing the solutions in the Pareto set. Subsequently, sets of non-dominated solutions would be obtained. This iterative process would make it possible to establish an ordering of the solution space according to the set which each solution is included in.




In multi-objective optimization problems, a suitable Pareto front gives more flexibility to select the solution with the best compromise between the optimization criteria.

The generation of the Pareto set has a high computational cost because the complexity of the underlying system makes the application of exact methods impossible. For this reason, in most cases, we find NP-complete problems, where, to obtain the optimum Pareto set, all solutions of the space of decision variables would have to be evaluated~\citep{zitzler2003performance}. The most common dealing with these optimization problems will be through the use of meta-heuristics which, despite not guaranteeing the identification of the real Pareto set, are capable of achieving a front close to the  optimum. Traditionally, evolutionary algorithms have been one of the tools used to solve these types of problems, with genetic algorithms being one of the most popular.

\subsection{Fog computing}
\label{sect_fogcomputing}

Fog computing is a computing paradigm that can be understood as an extension of cloud architecture, by including computational and storage capacity in the in-network devices,  located between the data center and the clients. The intermediate infrastructure is the extension of the cloud infrastructure creating a continuum between the cloud and the users.

In a fog computing architecture, the intermediate elements of the infrastructure, called fog nodes or fog devices, can store data or execute services/applications. Fog devices reduce the latency between users and application/data location. This improvement benefits those applications with critical response times. Fog devices are closer to users, and they are geographically distributed.

\begin{figure}
	
	\centering
	\includegraphics[width=\textwidth]{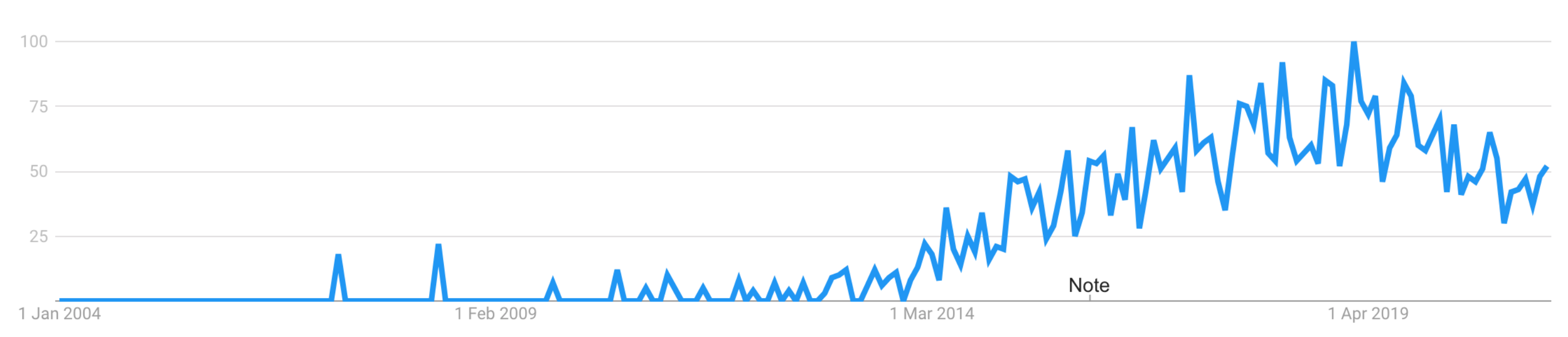}
	\caption{Analysis of search terms related to fog computing classified by year (Source: Google Trends, April 2021).}\label{trendsfog}
\end{figure}

\begin{figure}
	
	\centering
	\includegraphics[width=9cm]{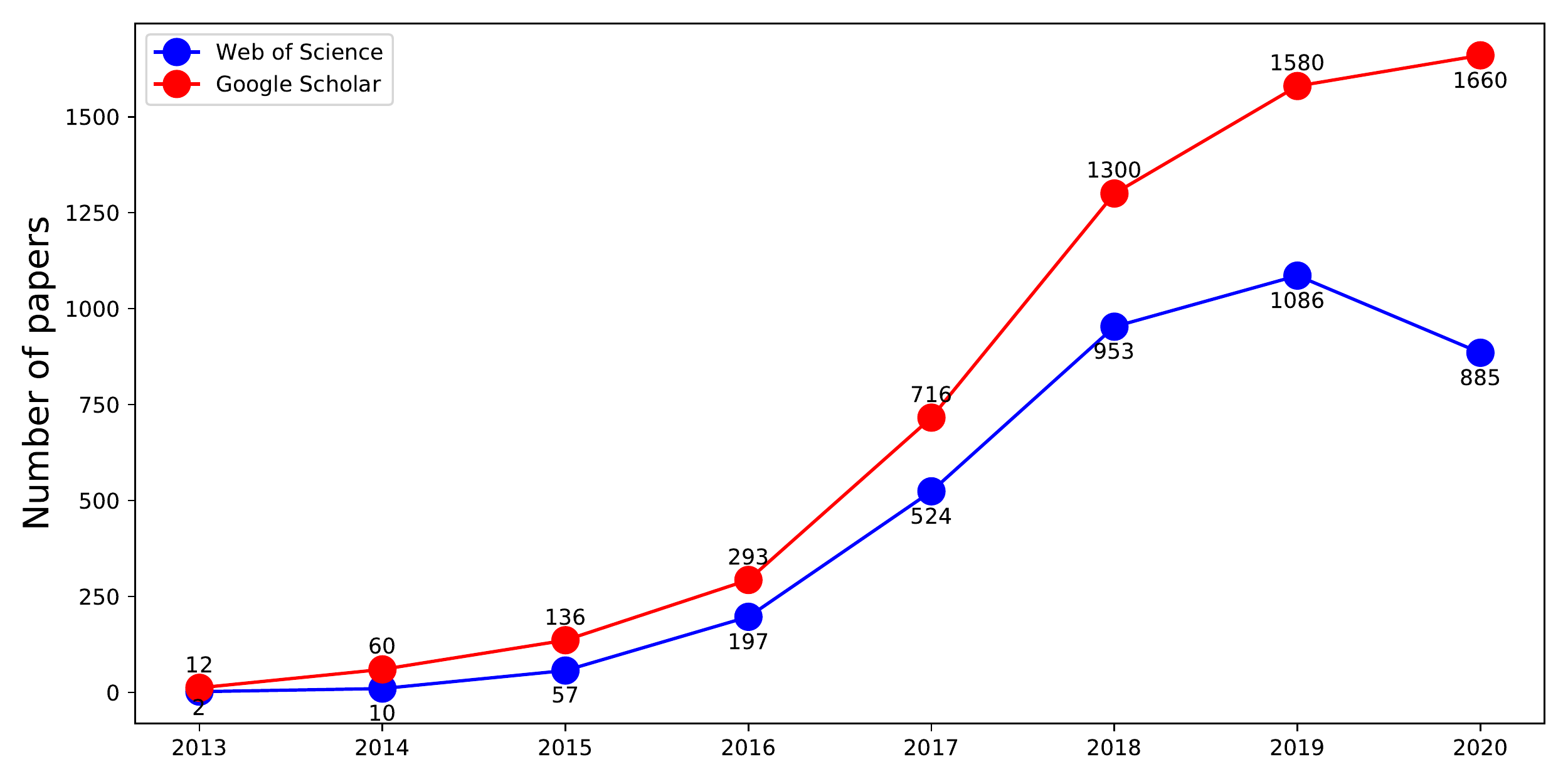}
	\caption{Number of scientific publications on fog computing grouped by publication year (Source: Web of Science and Google Scholar, March 2021).}\label{numpubsfog}
\end{figure}

For a clear vision of the popularity growth of this technology, Figure~\ref{trendsfog} shows the data obtained from Google Trends for the search term \textit{Fog computing}. Likewise, Figure~\ref{numpubsfog} shows the total number of publications per year that are indexed in Google Scholar and Web of Science and that contain the term \textit{Fog computing} in their titles or keywords. In both cases, we can see how the years 2013 and 2014 mark the beginning of the interest in fog technologies, two years after the first proposal of fog architecture \citep{bonomi2011connected,bonomi2012fog}.

The term Edge Computing is sometimes confused with fog computing, but there are clear differences between them \citep{nebbiolofogvsedge}. Fog computing is organized in different hierarchical levels that allow the dynamic configuration and orchestration of generic applications. On the contrary, Edge computing executes specific applications in a specific location limited to the devices at the peripheral level of the infrastructure. Edge computing systems are usually limited to computing and networking tasks, while fog environments also provide functionalities concerning data storage, control, and processing \citep{iorga2018fog}.

\begin{figure}
	\centering
	\includegraphics[width=10cm]{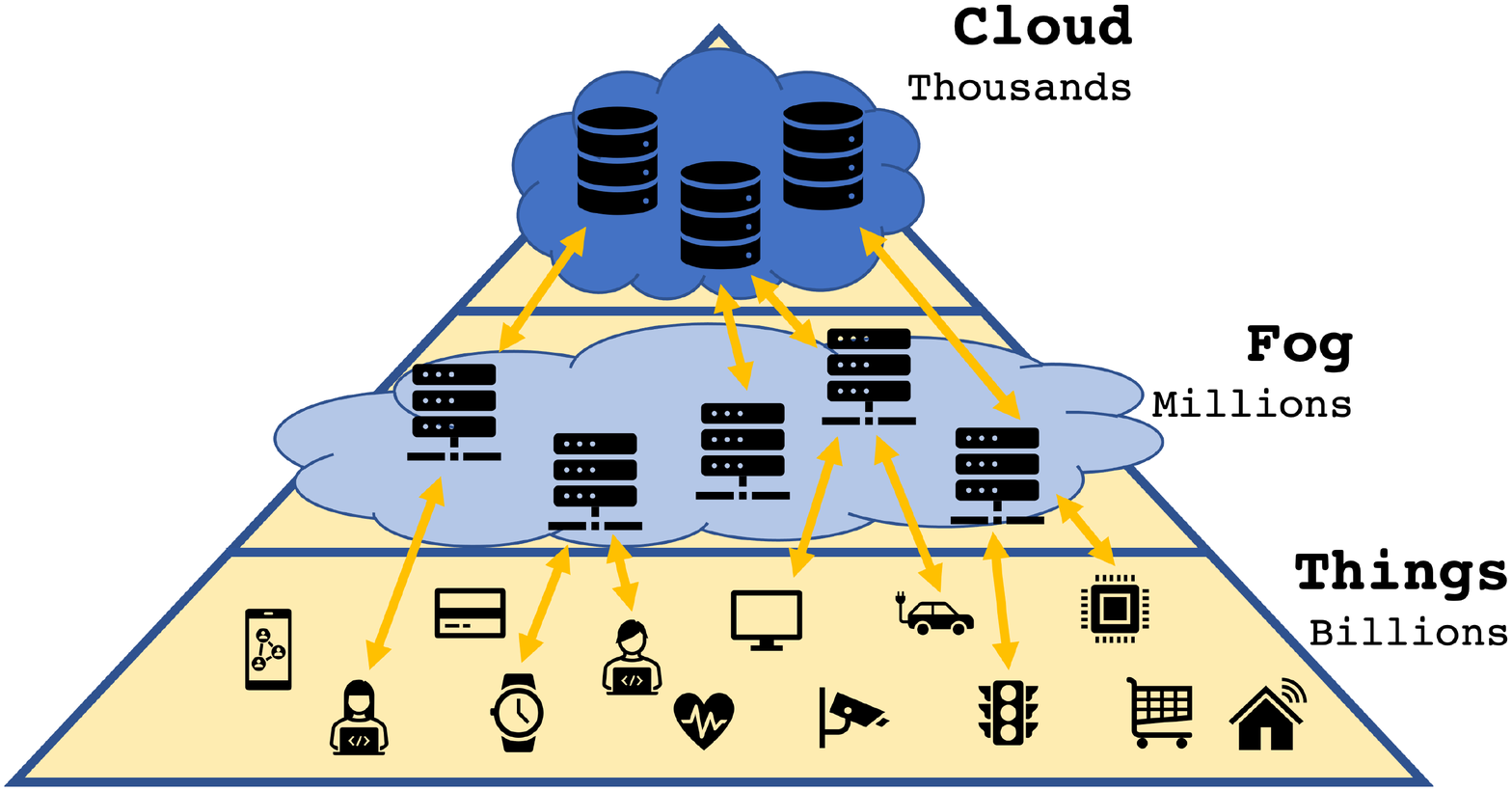}
	\caption{Basic architecture of fog computing.}\label{fogarchitecture}
\end{figure}

Fog infrastructures are usually split into three levels of abstraction (Figure~\ref{fogarchitecture}). At the highest level, we would find traditional cloud systems, responsible for providing data processing or management services. At the lowest level would be all users or the IoT devices (things) that generate data, consume it, or request the execution of services/applications. And the fog is the intermediate level, which is the extension of the cloud, offering the same type of services as the latter. Whether the services/data are in the nodes of the cloud system or on the fog level is transparent for the user.

In fog systems, as in any other distributed architecture, resource allocation, workload distribution, and infrastructure organization have a direct impact on the behavior of the system, applications, and users \citep{mann2017resource}. The quality of service will be directly related to these decision-making processes and will require optimization processes.

Although fog computing could be considered as a particular case of cloud computing, there are marked differences between them \citep{DBLP:conf/fwc/MayerGSR17,moysiadis2018}. The most important differences concern the scale, location, amount of resources, and the interconnection of the nodes (Table~\ref{fogvscloud}).

\begin{table}
	\caption{Differentiating features between fog computing and cloud computing architectures.}
	\label{fogvscloud}
	\centering
	\begin{tabular}{p{5.5cm}|p{3.5cm}p{3.5cm}}
		\toprule
		
		& \textbf{Cloud} & \textbf{Fog}\\
		\midrule
		Hardware resources & High & Limited \\ \midrule
		Hardware features & Homogeneous & Heterogeneous \\ \midrule
		Location & Limited and centralized & Geographically distributed \\
		& Static nodes & Static and mobile nodes \\ \midrule
		Network features  & Homogeneous & Heterogeneous \\
		& High speed, Low latency & Cases of high latency  \\ \midrule
		Scale / number of nodes& Thousands & Millions \\		
		\bottomrule
	\end{tabular}
\end{table}

These differentiating characteristics between cloud and fog make it necessary for a specific analysis of methodologies already adopted in cloud for the case of the fog, or the definition of new methodologies not previously applied. Therefore, with the definition of this new paradigm, a whole series of research lines are also opened with specific challenges to face. In particular, challenges related to high scale, limited resource availability, component failures, and heterogeneity of the infrastructure.

Although the use of GAs in fog optimization has already been explored in some studies \citep{guerrero2019evaluation, brogi2020place}, significant challenges remain open. In particular, and given the characteristics of fog systems, the use of parallel GAs and the hybrid combination with other meta-heuristics are perceived, a priory, as promising solutions. These expectations would be clarified by new contributions from the research works carried out in these lines.

\section{Methodology}
\label{sec_methodology}

This systematic literature research follows the common methodology used in the field of cloud or fog systems~\cite{OGUNDOYIN2021100937, barros2020scheduling}. The methodology establishes the need to define the search strategy, definition of the research questions, and procedures for paper selection. 

In this study, papers were searched from two databases: Google Scholar (GS) and Web of Science (WoS). GS was selected because of the large number of works it indexes and the short time between paper publication and GS indexing. This latter feature was considered to include the most recently published papers in the field. WoS was selected because of high quality papers and its advanced search options. 

Papers were included based on these criteria: (1) the objective should be to optimize resource management or service management, (2) the domain should be focused fog computing, and (3) the optimization should be performed using GAs.

The search term for GS was adapted to limit the results to the scope of fog computing. Many titles related to fog computing include only the term \textit{fog}. A search using this term would also include studies from other research areas, such as climatology. Consequently, the search term was defined to obtain all the articles that included the terms \textit{fog} in the title and \textit{ fog computing } in no less than one other part of the article. WoS includes a search option that limits the results to a specific research field. The previous adaptation was not required in the latter case. Table~\ref{searchstrings} lists the search terms for both databases.

\begin{table}[t!]
	\caption{Search terms for both databases.}
	\label{searchstrings}
	\centering
	\begin{tabular}{p{0.2\textwidth}|p{0.7\textwidth}}
		\toprule
		
		\textbf{Database} & \textbf{Search term}\\
		\midrule
		Web of Science & \texttt{TS = (fog computing) AND ALL = (optimization  OR scheduling  OR management  OR allocation  OR placement) AND ALL = (genetic)}\\
		Google Scholar & \texttt{'genetic algorithm' AND 'fog computing' AND optimization AND intitle:fog} \\
		\bottomrule
	\end{tabular}
\end{table}

Articles were searched at the beginning of April 2021, without limit on the publication year. The search returned a total of 77 articles from WoS and 475 from GS. These initial articles were successively filtered at different stages by reading the title, abstract, introduction, conclusions, and full article. After filtering, we finally included 70 articles in the systematic review.

The papers were analyzed to answer the following two research questions:
\begin{itemize}
	\item RQ1. Which resource optimization problems in fog environments have been addressed with GA? In terms of the optimization scope, what are open problems and research challenges?
	\item RQ2. What types of genetic design and GA have been implemented in fog resource optimization problems? What are the potential future research directions for GA design?
\end{itemize}

Consequently, the analysis of the papers is presented separately by answering both research questions (RQ1 in Section~\ref{sec:ambitodelsurvey} and RQ2 in Section~\ref{sec:surveyAG}). Inclusion and exclusion criteria were defined considering the research questions, that is, in terms of GA design and optimization scope.

Concerning the optimization scope, only articles that addressed the optimization of resource management or service management were included in the review. Papers that used fog architectures to optimize and improve processes (such as energy distribution, road traffic, and e-health applications\ldots) are beyond the scope of this systematic review. Papers in the domain of edge computing or mobile cloud computing were also excluded from the analysis. 

In relation to the design of GA, this systematic review only includes studies in which GA was a part of the paper’s contributions. By contrast, papers were excluded when they only used GA as an experimental control group. Additionally, the papers were only included if GA was used as an optimization algorithm \citep{app9061160}. Papers were excluded if they used GA as a complementary tool. For example, if GA was only used to assign weights to an artificial neural network \citep{KAUR2020101067}, the papers were not included in the analysis.

The filtering process included only articles that aligned with the previous inclusion and exclusion criteria. It also included articles from journals or conferences written in English. Our analysis did not include other literature reviews or surveys. It is important to mention that none of the 42 surveys obtained from searching in both databases cover the case considered in this analysis. Consequently, we can state that this is the first systematic review on the use of GAs for resource optimization in the field of fog computing. Additionally, the usefulness of this paper is justified because the number of found, filtered, and analyzed papers is very high. 

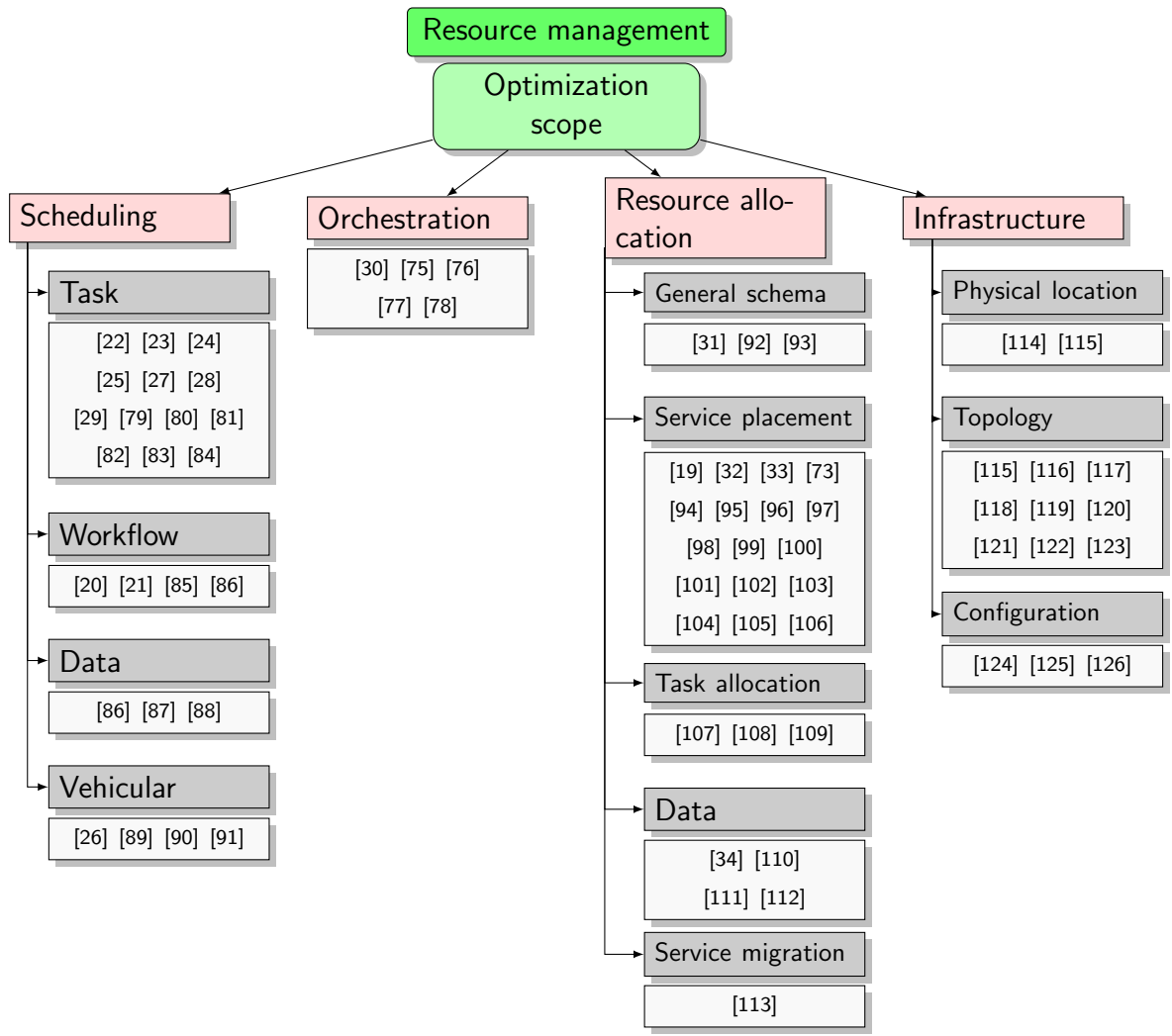
\begin{figure}
	\centering
	
\begin{tikzpicture}[
  level 1/.style={sibling distance=40mm},
  edge from parent/.style={->,draw},
  >=latex]

\node[root](root) {Resource management};
\node[level 2,below of = root] (c1) {Optimization scope}
    child{node[level 3] (c2) {Scheduling}} 
    child{node[level 3] (c3) {Orchestration}
        [level distance=0.95cm]
        child{node[level 5,rounded corners=0cm] (c31) {
        \scriptsize     
        \cite{REDDY2020102428} \cite{9300090} \cite{10.1007/978-981-15-0199-9_24}  \cite{9c434213a47d45b486ebc9c416d132d0} \cite{7867735}
        } edge from parent[draw=none] }
        edge from parent node{}
    }
    child{node[level 3] (c4) {Resource allocation}}
    child{node[level 3] (c5) {Infrastructure}}
    ;

%
%

\node [level 4, below of = c2, xshift=15pt] (c21) {Task}
    [level distance=1.45cm]
    child{node[level 5] (c211) {
    \scriptsize     
    \cite{AbbasiPK20} \cite{10.1016/j.future.2019.09.039} \cite{10.1145/3287921.3287984} \cite{app9091730} \cite{9140118} \cite{10.1002/dac.4652}  \cite{10.1007/s11277-017-5200-5}  \cite{10.3389/fphy.2020.00358} \cite{9233987}  \cite{8717643}  \cite{8359780} \cite{s19122783}  \cite{9308549} 
    } edge from parent[draw=none] };

\node [level 4, below of = c211,node distance=1.8cm] (c22) {Workflow}
    [level distance=0.7cm]
    child{node[level 5] (c221) {
    \scriptsize     
    \cite{8701449} \cite{DEMAIO2020171} \cite{10.23919/FRUCT.2017.8250177} \cite{Wang2018}
    } edge from parent[draw=none] };

\node [level 4, below of = c221] (c23) {Data}    
    [level distance=0.7cm]
    child{node[level 5] (c231) {
    \scriptsize     
    \cite{Wang2018} \cite{8085127} \cite{9306718}
    } edge from parent[draw=none] };

\node [level 4, below of = c231] (c24) {Vehicular}
    [level distance=0.7cm]
    child{node[level 5] (c241) {
    \scriptsize     
    \cite{LiZSS20} \cite{9045361} \cite{7545926} \cite{8761932}
    } edge from parent[draw=none] };

%
%
\node [level 4, below of = c4, xshift=15pt] (c41) {\footnotesize General schema}
    [level distance=0.7cm]
    child{node[level 5] (c411) {
    \scriptsize     
    \cite{8947936} \cite{8325027} \cite{s19061267} 
    } edge from parent[draw=none] };

\node [level 4, below of = c411] (c42) {\footnotesize Service placement}
    [level distance=1.75cm]
    child{node[level 5] (c421) {
    \scriptsize     
    \cite{guerrero2019evaluation} \cite{8676263} \cite{8929234} \cite{app9061160} \cite{9052663} \cite{9363254} \cite{9012671} \cite{8812204} \cite{9238236}  \cite{10.1145/3365871.3365892} \cite{7776569} \cite{9024660} \cite{NATESHA2021102972} \cite{9110359}  \cite{SkarlatNSBL17} \cite{SkarlatKRB018} \cite{VERBA201948} 
    } edge from parent[draw=none] };

\node [level 4, below of = c421,node distance=1.8cm] (c43) {\footnotesize Task allocation}
    [level distance=0.7cm]
    child{node[level 5] (c431) {
    \scriptsize     
    \cite{10.1007/978-3-030-33509-0_63} \cite{electronics9030474} \cite{8377192}
    } edge from parent[draw=none] };

\node [level 4, below of = c431] (c44) {Data}
    [level distance=0.95cm]
    child{node[level 5] (c441) {
    \scriptsize     
    \cite{a12100201} \cite{closer19canali} \cite{8735711} \cite{8630038}
    } edge from parent[draw=none] };

\node [level 4, below of = c441] (c45) {\footnotesize Service migration}
    [level distance=0.7cm]
    child{node[level 5] (c451) {
    \scriptsize     
    \cite{MartinKC20}
    } edge from parent[draw=none] };

%
%
\node [level 4, below of = c5, xshift=15pt] (c51) {\footnotesize Physical location}
    [level distance=0.7cm]
    child{node[level 5] (c511) {
    \scriptsize     
    \cite{10.1002/spe.2631} \cite{8611104}
    } edge from parent[draw=none] };
\node [level 4, below of = c511] (c52) {\footnotesize Topology}
    [level distance=1.225cm]
    child{node[level 5] (c521) {
    \scriptsize     
    \cite{8611104} \cite{HussainBA20} \cite{8339513} \cite{9117921} \cite{9220179} \cite{9022931} \cite{Vorobyev_2019} \cite{Vorobyev_2020} \cite{10.1007/978-3-030-62223-7_20} 
    } edge from parent[draw=none] };
\node [level 4, below of = c521,node distance=1.4cm] (c53) {\footnotesize Configuration}
    [level distance=0.7cm]
    child{node[level 5] (c531) {
    \scriptsize     
    \cite{BarikDMSM19} \cite{8422660} \cite{8254983}
    } edge from parent[draw=none] };


\foreach \value in {1,2,3,4}
   \draw[->] (c2.195) |- (c2\value.west);

\foreach \value in {1,2,3,4,5}
   \draw[->] (c4.195) |- (c4\value.west);
 
\foreach \value in {1,2,3}
   \draw[->] (c5.195) |- (c5\value.west);

\end{tikzpicture}

\caption{Taxonomy for the problem scope.}\label{fig:taxonomiaGR}
\end{figure}

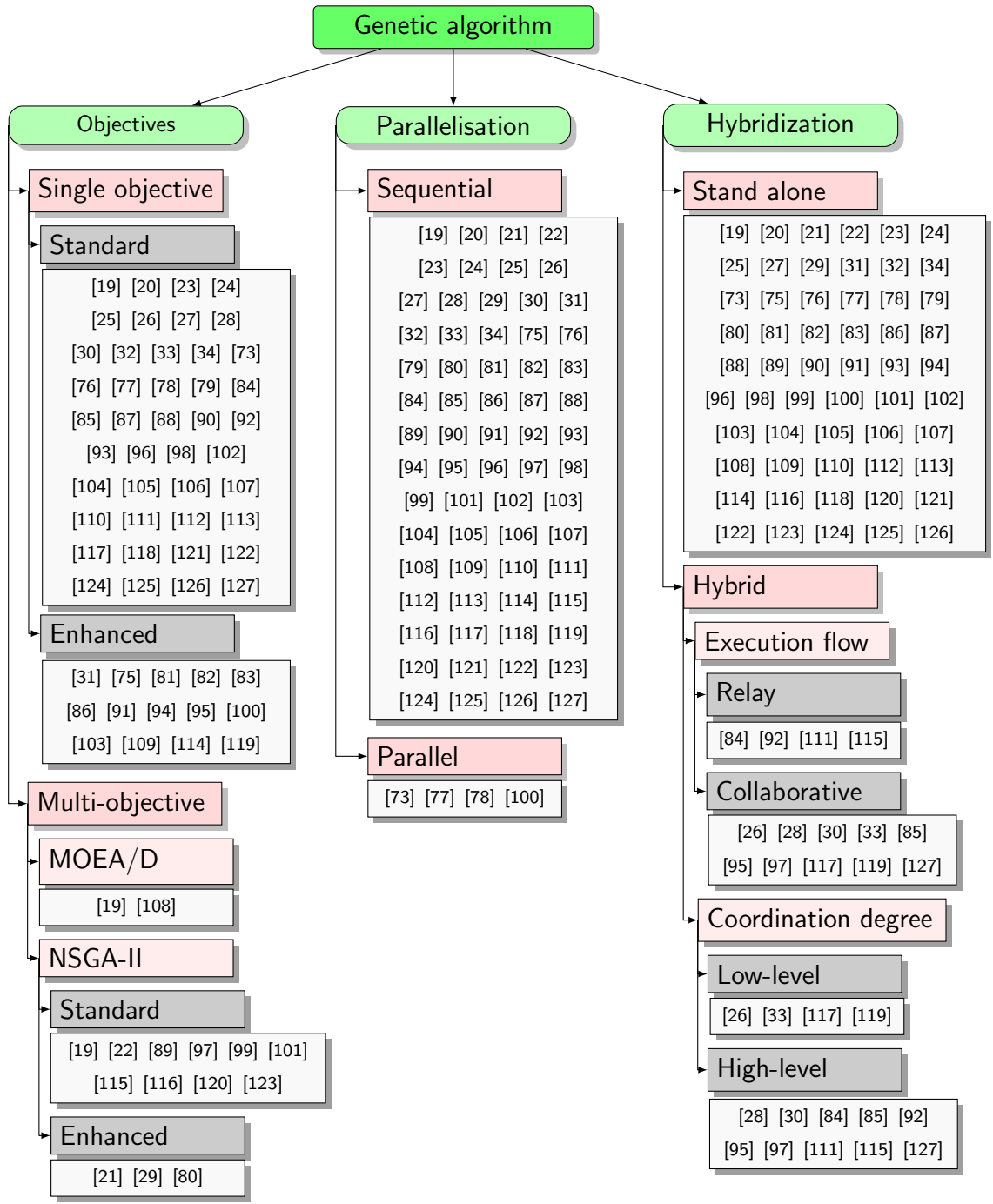
\begin{figure}
	\centering
\begin{tikzpicture}[
  level 1/.style={sibling distance=50mm},
  edge from parent/.style={->,draw},
  >=latex]

\node[root] {Genetic algorithm}
  child {node[level 2] (c1) {\footnotesize Objectives}}
  child {node[level 2] (c2) {Parallelisation}}
  child {node[level 2] (c3) {Hybridization}};


\begin{scope}[every node/.style={level 3}]
\node [below of = c1, xshift=0pt](c11) {Single objective};
\node [level 4,below =0.2cm of c11, xshift=5pt] (c111) {Standard}
    [level distance=2.95cm]
    child{node[level 5,xshift=12.5pt,text width=8.5em] (c1111) {
    \scriptsize     
\cite{guerrero2019evaluation}
\cite{8701449}
\cite{10.1016/j.future.2019.09.039}
\cite{10.1145/3287921.3287984}
\cite{app9091730}
\cite{LiZSS20}
\cite{9140118}
\cite{10.1002/dac.4652}
\cite{REDDY2020102428}
\cite{8676263}
\cite{8929234}
\cite{a12100201}
\cite{app9061160}
\cite{10.1007/978-981-15-0199-9_24}
\cite{9c434213a47d45b486ebc9c416d132d0}
\cite{7867735}
\cite{10.3389/fphy.2020.00358}
\cite{9308549}
\cite{10.23919/FRUCT.2017.8250177}
\cite{8085127}
\cite{9306718}
\cite{7545926}
\cite{8325027}
\cite{s19061267}
\cite{9012671}
\cite{9238236}
\cite{NATESHA2021102972}
\cite{SkarlatNSBL17}
\cite{SkarlatKRB018}
\cite{VERBA201948}
\cite{10.1007/978-3-030-33509-0_63}
\cite{closer19canali}
\cite{8735711}
\cite{8630038}
\cite{MartinKC20}
\cite{8339513}
\cite{9117921}
\cite{Vorobyev_2019}
\cite{Vorobyev_2020}
\cite{BarikDMSM19}
\cite{8422660}
\cite{8254983}
\cite{8311595}
    } edge from parent[draw=none] };

\node [level 4,below of = c1111,node distance=3.cm,xshift=-12.5pt] (c112) {Enhanced}    
    [level distance=1.2cm]
    child{node[level 5,text width=8.5em,xshift=12.5pt] (c1121) {
    \scriptsize     
    \cite{8947936} \cite{9300090} \cite{8717643} \cite{8359780} \cite{s19122783}  \cite{Wang2018} \cite{8761932} \cite{9052663} \cite{9363254} \cite{7776569} \cite{9110359}  \cite{8377192}  \cite{10.1002/spe.2631} \cite{9220179}
    } edge from parent[draw=none] };

\node [below of = c1121, xshift=-18pt,node distance=1.4cm] (c12) {Multi-objective};
\node [level 3a,below =0.2cm of c12, xshift=5pt] (c121) {MOEA/D}
    [level distance=0.7cm]
    child{node[level 5] (c1211) {
    \scriptsize     
    \cite{guerrero2019evaluation} \cite{electronics9030474}
    } edge from parent[draw=none] };
    
\node [level 3a,below =0.2cm of c1211] (c122) {NSGA-II};  
\node [level 4,below =0.2cm of c122, xshift=5pt] (c1221) {Standard}
    [level distance=0.9cm]
    child{node[level 5,text width=9.5em,xshift=17.5pt] (c12211) {
    \scriptsize     
    \cite{guerrero2019evaluation} \cite{AbbasiPK20} \cite{9045361} \cite{8812204} \cite{10.1145/3365871.3365892}  \cite{9024660}    \cite{8611104} \cite{HussainBA20} \cite{9022931} \cite{10.1007/978-3-030-62223-7_20}
    } edge from parent[draw=none] };
    
\node [level 4,below =0.2cm of c12211,node distance=1.2cm,xshift=-17.5pt] (c1222) {Enhanced}
    [level distance=0.65cm]
    child{node[level 5] (c12221) {
    \scriptsize     
     \cite{DEMAIO2020171} \cite{10.1007/s11277-017-5200-5} \cite{9233987}
    } edge from parent[draw=none] };

\node [below of = c2, xshift=5pt] (c21) {Sequential}    
    [level distance=4.25cm]
    child{node[level 5, xshift=12.5pt,text width=8.5em] (c211) {
    \scriptsize     
    \cite{guerrero2019evaluation}
\cite{8701449}
\cite{DEMAIO2020171}
\cite{AbbasiPK20}
\cite{10.1016/j.future.2019.09.039}
\cite{10.1145/3287921.3287984}
\cite{app9091730}
\cite{LiZSS20}
\cite{9140118}
\cite{10.1002/dac.4652}
\cite{10.1007/s11277-017-5200-5}
\cite{REDDY2020102428}
\cite{8947936}
\cite{8676263}
\cite{8929234}
\cite{a12100201}
\cite{9300090}
\cite{10.1007/978-981-15-0199-9_24}
\cite{10.3389/fphy.2020.00358}
\cite{9233987}
\cite{8717643}
\cite{8359780}
\cite{s19122783}
\cite{9308549}
\cite{10.23919/FRUCT.2017.8250177}
\cite{Wang2018}
\cite{8085127}
\cite{9306718}
\cite{9045361}
\cite{7545926}
\cite{8761932}
\cite{8325027}
\cite{s19061267}
\cite{9052663} 
\cite{9363254}
\cite{9012671}
\cite{8812204}
\cite{9238236}
\cite{10.1145/3365871.3365892}
\cite{9024660}
\cite{NATESHA2021102972}
\cite{9110359}
\cite{SkarlatNSBL17}
\cite{SkarlatKRB018}
\cite{VERBA201948}
\cite{10.1007/978-3-030-33509-0_63}
\cite{electronics9030474}
\cite{8377192}
\cite{closer19canali}
\cite{8735711}
\cite{8630038}
\cite{MartinKC20}
\cite{10.1002/spe.2631}
\cite{8611104}
\cite{HussainBA20}
\cite{8339513}
\cite{9117921}
\cite{9220179}
\cite{9022931}
\cite{Vorobyev_2019}
\cite{Vorobyev_2020}
\cite{10.1007/978-3-030-62223-7_20}
\cite{BarikDMSM19}
\cite{8422660}
\cite{8254983}
\cite{8311595}
    } edge from parent[draw=none] };
    
\node [below of = c211,node distance=4.4cm, xshift=-12.5pt] (c22) {Parallel}
    [level distance=0.65cm]
    child{node[level 5] (c221) {
    \scriptsize     
    \cite{app9061160} \cite{9c434213a47d45b486ebc9c416d132d0} \cite{7867735} \cite{7776569} 
    } edge from parent[draw=none] };

\node [below of = c3, xshift=0pt](c31a) {Stand alone}
    [level distance=2.95cm]
    child{node[level 5, xshift=23.5pt,text width=10.5em] (c31a1) {
    \scriptsize     
\cite{guerrero2019evaluation}
\cite{8701449}
\cite{DEMAIO2020171}
\cite{AbbasiPK20}
\cite{10.1016/j.future.2019.09.039}
\cite{10.1145/3287921.3287984}
\cite{app9091730}
\cite{9140118}
\cite{10.1007/s11277-017-5200-5}
\cite{8947936}
\cite{8676263}
\cite{a12100201}
\cite{app9061160}
\cite{9300090}
\cite{10.1007/978-981-15-0199-9_24}
\cite{9c434213a47d45b486ebc9c416d132d0}
\cite{7867735}
\cite{10.3389/fphy.2020.00358}
\cite{9233987}
\cite{8717643}
\cite{8359780}
\cite{s19122783}
\cite{Wang2018}
\cite{8085127}
\cite{9306718}
\cite{9045361}
\cite{7545926}
\cite{8761932}
\cite{s19061267}
\cite{9052663} 
\cite{9012671}
\cite{9238236}
\cite{10.1145/3365871.3365892}
\cite{7776569}
\cite{9024660}
\cite{NATESHA2021102972}
\cite{9110359}
\cite{SkarlatNSBL17}
\cite{SkarlatKRB018}
\cite{VERBA201948}
\cite{10.1007/978-3-030-33509-0_63}
\cite{electronics9030474}
\cite{8377192}
\cite{closer19canali}
\cite{8630038}
\cite{MartinKC20}
\cite{10.1002/spe.2631}
\cite{HussainBA20}
\cite{9117921}
\cite{9022931}
\cite{Vorobyev_2019}
\cite{Vorobyev_2020}
\cite{10.1007/978-3-030-62223-7_20}
\cite{BarikDMSM19}
\cite{8422660}
\cite{8254983}
    } edge from parent[draw=none] };

 
\node [below=0.2cm of c31a1, xshift=-23.5pt](c31b) {Hybrid};
\node [level 3a,below =0.2cm of c31b, xshift=5pt] (c31) {Execution flow};
\node [level 4,below =0.2cm of c31, xshift=5pt] (c311) {Relay}
    [level distance=0.7cm]
    child{node[level 5] (c3111) {
    \scriptsize     
\cite{9308549} \citep{8325027} \citep{8735711} \citep{8611104}
    } edge from parent[draw=none] };
    
\node [level 4,below =0.2cm of c3111 ] (c312) {Collaborative}
    [level distance=0.9cm]
    child{node[level 5,text width=8.5em, xshift=12.5pt] (c3121) {
    \scriptsize     
\cite{LiZSS20} \cite{10.1002/dac.4652} \cite{REDDY2020102428} \cite{8929234} \cite{10.23919/FRUCT.2017.8250177} \cite{9363254} \cite{8812204} \cite{8339513} \cite{9220179} \cite{8311595}
    } edge from parent[draw=none] };

\node [level 3a,below =0.2cm of c3121, xshift=-4.5pt,node distance=1.7cm,text width=8.5em] (c32) {Coordination degree};
\node [level 4,below =0.2cm of c32, xshift=-7.5pt] (c321) {Low-level}
    [level distance=0.65cm]
    child{node[level 5,xshift=1.0pt] (c3211) {
    \scriptsize     
\cite{LiZSS20} \cite{8929234} \cite{8339513} \cite{9220179}
    } edge from parent[draw=none] };
    
\node [level 4,below =0.2cm of c3211,xshift=-1.0pt] (c322) {High-level}
    [level distance=0.98cm]
    child{node[level 5,text width=8.5em,xshift=12.50pt] (c3221) {
    \scriptsize     
\cite{10.1002/dac.4652} \cite{REDDY2020102428} \cite{9308549} \cite{10.23919/FRUCT.2017.8250177} \cite{8325027} \cite{9363254} \cite{8812204} \cite{8735711} \cite{8611104} \cite{8311595}
    } edge from parent[draw=none] };

\end{scope}

\foreach \value in {1,2}
  \draw[->] (c1.west) |- (c1\value.west);

\foreach \value in {1,2}
  \draw[->] (c2.west) |- (c2\value.west);

\foreach \value in {1,2}
  \draw[->] (c31b.west) |- (c3\value.west);

\draw[->] (c3.west) |- (c31a.west) ;
\draw[->] (c3.west) |- (c31b.west) ;

\draw[->] (c11.west) |- (c111.west) ;
\draw[->] (c11.west) |- (c112.west) ;

\draw[->] (c12.west) |- (c121.west) ;
\draw[->] (c12.west) |- (c122.west) ;

\draw[->] (c122.west) |- (c1221.west) ;
\draw[->] (c122.west) |- (c1222.west) ;

\draw[->] (c31.west) |- (c311.west) ;
\draw[->] (c31.west) |- (c312.west) ;

\draw[->] (c32.west) |- (c321.west) ;
\draw[->] (c32.west) |- (c322.west) ;

\end{tikzpicture}
\caption{ Taxonomy for the genetic algorithm.}\label{fig:taxonomiaAG}
\end{figure}

The analysis of the 70 selected articles was divided into two dimensions. First, they were organized according to the domain of the optimization problem. Figure~\ref{fig:taxonomiaGR} shows the taxonomy of the different types of optimization problems. Subsequently, the papers were analyzed according to the GA design. Figure~\ref{fig:taxonomiaAG} shows the taxonomy of the GA design.



\section{Scope of the optimization}
\label{sec:ambitodelsurvey}


Optimization and resource management problems are usually classified into scheduling, allocation, load balancing, and provisioning. It is important to note that the boundary among different groups is not always clearly defined~\citep{10.1007/s10115-016-0951-y, MANVI2014424}, and some of these concepts are sometimes used interchangeably.

We adapted this basic optimization-domain taxonomy to our case, which emerged from the papers included in the systematic review. Figure~\ref{fig:taxonomiaGR} shows the taxonomy used to group the papers. We used the optimization field indicated in the paper or, if not indicated, we classify it as the most appropriate group. Tables~\ref{tab:surveysegunPlanificacion}, \ref{tab:surveysegunOrquestacion}, \ref{tab:surveysegunAsignacion}, and~\ref{tab:surveysegunInfraestructura} show the general organization of the works according to the optimization domain, and they also include an additional column with an acronym that indicates the GA design for each case.

\subsection{Scheduling}

\begin{table}[t!]
	\caption{Papers within the scope of scheduling optimization.}
	\label{tab:surveysegunPlanificacion}
	\centering
		\tabcolsep=0.07cm
		\footnotesize
	\begin{tabular}{p{6cm}|ccccl}\toprule
		&	\rotatebox{90}{Workflow scheduling}	&	\rotatebox{90}{Task scheduling}	&	\rotatebox{90}{Data workflow}& \rotatebox{90}{Vehicular}&	Algorithm$^a$\\ \midrule
		
		\citeauthor{8701449}~\cite{8701449}	&	$\bullet$	&		&		&&	eGA\\
		
		\citeauthor{DEMAIO2020171}~\cite{DEMAIO2020171}	&	$\bullet$	&		&		&&	eNSGA2\\
		
		\citeauthor{10.23919/FRUCT.2017.8250177}~\cite{10.23919/FRUCT.2017.8250177} &$\bullet$&&&&Hyb.\\								
		
		﻿﻿\citeauthor{Wang2018}~\cite{Wang2018} 	&	$\bullet$	&		&	$\bullet$	&&	eGA\\

		\citeauthor{AbbasiPK20}~\cite{AbbasiPK20}	&		&	$\bullet$	&		&&	NSGA2\\
		
		\citeauthor{10.1016/j.future.2019.09.039}~\cite{10.1016/j.future.2019.09.039}&&$\bullet$&&&GA\\								
		
		﻿\citeauthor{10.3389/fphy.2020.00358}~\cite{10.3389/fphy.2020.00358}	&		&	$\bullet$	&		&&	GA\\
		
		﻿\citeauthor{9233987}~\cite{9233987}	&		&	$\bullet$	&		&&	eNSGA2\\
		
		﻿\citeauthor{10.1145/3287921.3287984}~\cite{10.1145/3287921.3287984,app9091730} 	&		&	$\bullet$	&		&&	GA\\

		﻿\citeauthor{8717643}~\cite{8717643}	&		&	$\bullet$	&		&&	eGA\\

		\citeauthor{8359780}~\cite{8359780}	&		&	$\bullet$	&		&&	eGA\\
		
		﻿\citeauthor{s19122783}~\cite{s19122783} 	&		&	$\bullet$	&		&&	eGA\\

		\citeauthor{9140118}~\cite{9140118}	&		&	$\bullet$	&		&&	GA\\
		
		﻿\citeauthor{10.1002/dac.4652}~\cite{10.1002/dac.4652}	&		&	$\bullet$	&		&&	Hyb.\\
		
		﻿\citeauthor{9308549}~\cite{9308549}	&		&	$\bullet$	&		&&	Hyb.\\
		
		﻿\citeauthor{10.1007/s11277-017-5200-5}~\cite{10.1007/s11277-017-5200-5}	&		&	$\bullet$	&		&&	eNSGA2\\

		﻿\citeauthor{8085127}~\cite{8085127} 	&		&		&	$\bullet$	&&	GA\\
		\citeauthor{9306718}~\cite{9306718} 	&		&		&	$\bullet$	&&	GA\\
		
\citeauthor{8761932}~\cite{8761932}&&&&$\bullet$&eGA\\
\citeauthor{LiZSS20}~\cite{LiZSS20}&&&&$\bullet$&Hyb.\\
\citeauthor{9045361}~\cite{9045361}&&&&$\bullet$&NSGA2\\
\citeauthor{7545926}~\cite{7545926}&&&&$\bullet$&GA\\

		\bottomrule
	\end{tabular}
	
	\begin{flushleft}
		\scriptsize$^a$ GA: classic genetic algorithm, eGA: enhanced GA, eNSGA2: enhanced NSGA-II, Hyb.: hybrid GA, Par.: parallel GA 
	\end{flushleft}
\end{table}

In scheduling optimization, a job $J$ is defined as a set of tasks $T=\{T_1,\ldots,T_n\}$. The scheduling algorithm determines the execution timing of these tasks. A job can be defined as a set of independent tasks (task scheduling) or by considering the dependencies between tasks (workflow scheduling)~\citep{10.1145/3325097}. Task execution planning is also constrained by the resources available in the computing devices $R=\{R_1,\ldots,R_m\}$.

In the case of workflow scheduling, execution dependencies are defined as a directed acyclic graph (DAG) $G(T,P)$, where nodes are tasks, and edges are temporal dependencies or precedence constraints $P$ between the tasks. Consequently, a graph $J=G(T,P)$ represents the jobs in the system.

Scheduling optimization includes finding both tasks to be allocated $T$ to available computing resources, $f:T \rightarrow R$, as well as the temporal planning of task executions, $g: T \rightarrow t_{clock}$, in such a way that the timing constraints are fulfilled with the finite available computational resources.

Makespan is one of the common metrics studied in such problems. Makespan refers to the amount of time elapsed between the start of the first task and the end of the last task. Another commonly studied metric is the resource usage execution cost.

The following subsections include papers within the scope of scheduling optimization (Table~\ref{tab:surveysegunPlanificacion}).

\subsubsection{Task scheduling}

As previously mentioned, in the particular case of independent task planning, the planning process does not consider dependencies between tasks $T=\{T_1,\ldots,T_n\}$ of job $J$.

\citeauthor{9308549}~\cite{9308549} studied the task scheduling of resources defined as virtual machines hosted in cloud and fog layers nodes. They used several algorithms based on the evolution of the population. Specifically, they proposed optimizing the execution time, execution cost, and energy consumption of the entire task. They used a single-objective optimization function calculated through a weighted transformation of these three criteria. They presented a hybrid GA that generated the first set of solutions using a classical GA, and a particle swarm optimization (PSO) that improves the solutions in that initial set.

\citeauthor{10.1145/3287921.3287984}~\cite{10.1145/3287921.3287984} and \citeauthor{10.3389/fphy.2020.00358}~\cite{10.3389/fphy.2020.00358} similarly proposed a classic GA for the scheduling of independent and parallel executed tasks, without considering synchronization or communication between them. Tasks are executed either on fog devices or clouds. First, \citeauthor{10.1145/3287921.3287984}~\cite{10.1145/3287921.3287984} defined a weight-based utility function that relates execution time and execution cost and, in a successive study \citep{app9091730}, compared it with a PSO algorithm. By contrast, \citeauthor{10.3389/fphy.2020.00358}~\cite{10.3389/fphy.2020.00358} only optimized the energy consumption.

\citeauthor{9140118}~\cite{9140118} also proposed using a classic GA. The main difference is that this proposal rejects task execution if the deadline is not met. The objective function is the relationship between the monetary cost and the percentage of tasks not executed within their execution deadlines.

\citeauthor{10.1016/j.future.2019.09.039}~\cite{10.1016/j.future.2019.09.039} proposed a classical GA with a fitness function calculated as the weighted average of the execution latency. Thus, the function assigns a greater weight to higher-priority tasks.

\citeauthor{8359780}~\cite{8359780} proposed to adapt crossover and mutation probabilities throughout the execution of a GA. The probabilities are reduced as the fitness function is improved. The fitness function considers execution time and communication cost.

\citeauthor{10.1007/s11277-017-5200-5}~\cite{10.1007/s11277-017-5200-5} proposed a two-tier solution for task planning. They assumed that the fog nodes were organized into groups called fog clusters. First, the scheduling algorithm selects the fog cluster to execute the task, and then it chooses a fog device within the fog cluster. The optimization criteria considered the execution time and task reliability. To achieve this, they proposed the use of the non-dominated sorting genetic algorithm-II (NSGA-II) algorithm, with an improvement in the crowding distance calculation function.

\citeauthor{9233987}~\cite{9233987} addressed the task scheduling problem using a modified NSGA-II algorithm with discrete encoding rather than a mapping structure between compute nodes and tasks. This encoding helps to save computing resources, particularly concerning the storage and processing of solutions.

Similarly, \citeauthor{8377192}~\cite{8377192} proposed the use of a GA with real coding. The problem studied was focused on an industrial IoT environment, and the objective was to achieve a schedule that equally distributes the subtasks among all fog devices.

\citeauthor{AbbasiPK20}~\cite{AbbasiPK20} also considered a NSGA-II algorithm to study the case for multiple cloud providers, in which tasks can be executed in the fog devices or any cloud provider. The NSGA-II algorithm must balance the power consumption with the execution time. In addition, the optimization problem imposed restrictions on the load balancing, resource capacity, total power consumption, and execution time of the entire system.

\citeauthor{s19122783}~\cite{s19122783} modified the crossover operation to improve the diversity of results. Thus, the crossover operation generates three solutions rather than two. Typically, two solutions are generated by combining parts of the parent chromosomes. The third solution is obtained as the average of the genes in the parent solutions.

\citeauthor{8717643}~\cite{8717643} proposed to incorporate a penalty function in the GA design. This penalty function reduces the fitness of solutions that do not meet the constraints of the problem.

\citeauthor{10.1002/dac.4652}~\cite{10.1002/dac.4652} proposed to solve the task scheduling problem in fog environments by combining a GA and ant colony optimization (ACO). They designed a coordinated execution of both algorithms by exchanging the best solutions between them, with the objective of minimizing the total execution time and energy consumption of the fog system.

\subsubsection{Workflow scheduling}

In workflow scheduling, the execution workflow is modeled by a DAG, $G(T,P)$, where nodes represent tasks, $T=\{T_1,\ldots,T_n\}$, and edges correspond to temporal dependencies or precedence constraints between them, $P$. Therefore, the scheduler considers these precedence relationships, and a task execution cannot begin until all preceding tasks are completed. Within this problem scope, we found only three works that used GA.

\citeauthor{DEMAIO2020171}~\cite{DEMAIO2020171} handled the problem of scheduling highly complex workflows. They considered dependencies between tasks and optimized makespan, execution cost, and reliability. They designed an NSGA-II algorithm modified with a simulated binary crossover operation (SBX) and a polynomial mutation. In addition, they considered a second vector in the representation of the solutions. The vector represented the dependencies and synchronization points between tasks. The evaluation of the solutions also considered the infrastructure topology and its influence on the workflow.

\citeauthor{10.23919/FRUCT.2017.8250177}~\cite{10.23919/FRUCT.2017.8250177} addressed the problem of scheduling complex execution workflows by using a hyper-heuristic algorithm. Hyper-heuristic algorithms combine a set of heuristics to compensate for the disadvantages of each heuristic. More specifically, this work proposed to combine GA, PSO, ACO, and simulated annealing (SA). The test-and-select strategy chooses the most appropriate solutions among the four algorithms. The goal is to optimize the energy consumption, runtime, network usage, and total cost. In a recent study, the same authors addressed additional considerations related to security \citep{8311595}.

\subsubsection{Data processing scheduling}

Fog architectures address the reduction in the amount of data transmitted between the sensors and cloud. Data generators and data processors (or preprocessors) are located closer to each other. Consequently, network usage is considerably reduced because of the lesser amount of transmitted data owing to its preprocessing. Therefore, specific studies are required for partitioning and scheduling of the processing of these data \citep{DIASDEASSUNCAO20181}.

\citeauthor{8085127}~\cite{8085127} superficially addressed the problem of sensor data partitioning and the plan to process it. Without including the exact details, they proposed a GA to solve the scheduling problem and a Dirichlet distribution to generate a set of random vectors for the data partitioning problem.

\citeauthor{9306718}~\cite{9306718} proposed to optimize the energy consumption, service latency, and total execution cost of the tasks that preprocess data in the fog nodes. To achieve this, they evaluated a variety of algorithms, including a classic GA. These algorithms plan the tasks, considering both the cloud level and fog devices.

\citeauthor{8701449}~\cite{8701449} studied the workflow scheduling for data stream applications that process a continuous flow of data. They divided the optimization process into two steps. In the first step, resources are allocated through a gradient-based adaptive process. In the second step, task scheduling is defined using a GA that applies the concept of elitism to the evolution of generations.

\citeauthor{Wang2018}~\cite{Wang2018} targeted their proposal on the scope of large-scale workflow problems. Specifically, they focused on addressing the scheduling of large volumes of astronomical data (a generation rate of 200 GB per day with a total size of more than 125 TB). They proposed the use of a classic GA that incorporates a local search operator to speed up the optimization process.

\subsubsection{Scheduling in vehicular domains}

Fog computing has resulted in a successful paradigm and in the emergence of new extensions, such as vehicular fog computing (VFC). VFC conceives vehicles as a collaborative architecture in which users incorporate computing and communication capabilities \citep{7415983}. Through the dynamic aggregation of vehicle resources, it is possible to improve the quality of service offered by the fog architecture, which is characterized by being highly dynamic.

\citeauthor{9045361}~\cite{9045361} handled the scheduling of tasks in VFC environments. They proposed the use of an NSGA-II algorithm that selects vehicles to execute tasks, or they are assigned to other devices on the network. The goal is to balance the minimization of resource usage and maximize the number of tasks executed in the vehicles.

\citeauthor{7545926}~\cite{7545926} handled a similar optimization problem in which vehicles correspond to public transport buses. These vehicles include the processing capacity to offer computer services to bus passengers. To achieve this, they proposed the use of a classical GA to minimize the cost of data transmission, limited by the constraints of satisfying a minimum quality of service.

\citeauthor{LiZSS20}~\cite{LiZSS20} addressed the problem of task offloading and its partition into independent subtasks. They proposed a hybrid algorithm that combines the characteristics of a GA with a beetle antennae search (BAS) algorithm. To achieve this, they incorporated a new step into the structure of classical GA, which evaluates the direction vectors of BAS. Thus, the solutions evolve using a direction vector that offers the best result.

Finally, \citeauthor{8761932}~\cite{8761932} studied task scheduling in vehicular environments for swarms of drones. The main objective was to reduce energy consumption while minimizing latency and availability. For this, they proposed the use of a GA with real coding.

\subsection{Service orchestration}

\begin{table}[t!]
	\caption{Papers within the scope of optimizing service orchestration.}
	\label{tab:surveysegunOrquestacion}
	\centering
	\footnotesize
	\begin{tabular}{lcl}\toprule
		& Orchestration & Algorithm$^a$\\ \midrule
		\citeauthor{9300090}~\cite{9300090} &	$\bullet$	&	eGA \\
		\citeauthor{10.1007/978-981-15-0199-9_24}~\cite{10.1007/978-981-15-0199-9_24}	&	$\bullet$	&	GA \\
		\citeauthor{REDDY2020102428}~\cite{REDDY2020102428} 	&	$\bullet$	&	Hyb. \\
		\citeauthor{9c434213a47d45b486ebc9c416d132d0}~\cite{9c434213a47d45b486ebc9c416d132d0,7867735} 	&	$\bullet$	&	Par.\\
		\bottomrule
	\end{tabular}
	
	\begin{flushleft}
		\scriptsize$^a$ GA: classic genetic algorithm, eGA: enhanced GA, eNSGA2: enhanced NSGA-II, Hyb.: hybrid GA, Par.: parallel GA 
	\end{flushleft}
\end{table}

In the domain of this survey, the term orchestration refers to the process of selecting, integrating, and composing services to satisfy user needs. This process considers resource availability and network management, and it covers all levels in the infrastructure (cloud-fog-users) \citep{8292756}. Cloud orchestration strategies cannot be applied without a previous evaluation in fog environments because of their different features, such as dynamism, scale, heterogeneity, and geographical distribution \citep{VAQUERO201920,Velasquez2018}. Table~\ref{tab:surveysegunOrquestacion} lists the papers within this optimization domain.

\citeauthor{10.1007/978-981-15-0199-9_24}~\cite{10.1007/978-981-15-0199-9_24} addressed the problem of selecting and composing of fog services. They assumed a fog system with multiple instances of services that are necessary to respond to user requests. Users request jobs that require the composition of a set of services. The services have multiple instances, and the algorithm needs to select one among them. They proposed a classic GA with a weighted fitness function that considers execution time, economic cost, availability, and reliability as the optimization criteria.

\citeauthor{9c434213a47d45b486ebc9c416d132d0}~\cite{9c434213a47d45b486ebc9c416d132d0} and \citeauthor{7867735}~\cite{7867735} handled service orchestration and workflow composition in scenarios with a high number of services. They proposed the use of a parallel version of the GA using Spark. The goal was to improve the security and performance.

\citeauthor{9300090}~\cite{9300090} also addressed the problem of service composition to improve seven metrics concerning the quality of service. To achieve this, they proposed using a multi-population GA. These GAs subdivide the population into different groups based on the values of the fitness function. Fitness was calculated through a weighted transformation of the seven metrics considered.

\subsection{Resource allocation}

\begin{table}[t!]
	\caption{Papers within the scope of resource allocation.}
	\label{tab:surveysegunAsignacion}
	\centering
		\tabcolsep=0.07cm
		\footnotesize
	\begin{tabular}{p{5cm}|cccccl}\toprule
		
		&	\rotatebox{90}{\parbox{2cm}{General}}	&	\rotatebox{90}{\parbox{2cm}{Service \mbox{placement}}}	&	\rotatebox{90}{\parbox{2cm}{Tasks \mbox{allocation}}}	&	\rotatebox{90}{\parbox{2cm}{Data allocation}}	&	\rotatebox{90}{\parbox{2cm}{Migration}}	&	Algorithm$^a$ \\

		\midrule﻿
		
		﻿\citeauthor{8325027}~\cite{8325027}	&	$\bullet$	&		&		&		&		&	Hyb. \\
		\citeauthor{s19061267}~\cite{s19061267} 	&	$\bullet$	&		&		&		&		&	GA \\
		﻿
		﻿\citeauthor{8947936}~\cite{8947936}	&	$\bullet$	&		&		&		&		&	eGA \\

		\citeauthor{9052663}~\cite{9052663}	&		&	$\bullet$	&		&		&		&	eGA \\
		
		﻿\citeauthor{9363254}~\cite{9363254}	&		&	$\bullet$	&		&		&		&	Hyb. \\
		
		﻿\citeauthor{9012671}~\cite{9012671}	&		&	$\bullet$	&		&		&		&	GA \\
		
		﻿\citeauthor{8812204}~\cite{8812204} 	&		&	$\bullet$	&		&		&		&	Hyb. \\
		
		﻿\citeauthor{9238236}~\cite{9238236} 	&		&	$\bullet$	&		&		&		&	GA \\
		
		﻿﻿\citeauthor{guerrero2019evaluation}~\cite{guerrero2019evaluation}	&		&	$\bullet$	&		&		&		&	GA NSGA2 MOEA/D \\
		
		﻿\citeauthor{10.1145/3365871.3365892}~\cite{10.1145/3365871.3365892} 	&		&	$\bullet$	&		&		&		&	NSGA2 \\
		
		﻿\citeauthor{7776569}~\cite{7776569}	&		&	$\bullet$	&		&		&		&	Par. \\
		
		﻿\citeauthor{9024660}~\cite{9024660} 	&		&	$\bullet$	&		&		&		&	NSGA2 \\
		
		﻿\citeauthor{NATESHA2021102972}~\cite{NATESHA2021102972} 	&		&	$\bullet$	&		&		&		&	eGA \\
		
		﻿\citeauthor{9110359}~\cite{9110359}	&		&	$\bullet$	&		&		&		&	eGA \\
		
		﻿\citeauthor{app9061160}~\cite{app9061160}	&		&	$\bullet$	&		&		&		&	Par. \\
		
		﻿\citeauthor{SkarlatNSBL17}~\cite{SkarlatNSBL17,SkarlatKRB018} 	&		&	$\bullet$	&		&		&		&	GA \\
		
		﻿\citeauthor{VERBA201948}~\cite{VERBA201948}	&		&	$\bullet$	&		&		&		&	GA \\
		
		﻿\citeauthor{8676263}~\cite{8676263}	&		&	$\bullet$	&		&		&		&	GA \\
		
		﻿\citeauthor{8929234}~\cite{8929234} 	&		&	$\bullet$	&		&		&		&	Hyb. \\

		\citeauthor{10.1007/978-3-030-33509-0_63}~\cite{10.1007/978-3-030-33509-0_63} 	&		&		&	$\bullet$	&		&		&	GA \\
		﻿\citeauthor{electronics9030474}~\cite{electronics9030474}	&		&		&	$\bullet$	&		&		&	MOEA/D \\
		﻿﻿\citeauthor{8377192}~\cite{8377192}	&		&		&	$\bullet$	&		&		&	eGA \\
		
		﻿\citeauthor{closer19canali}~\cite{a12100201,closer19canali} 	&		&		&		&	$\bullet$	&		&	GA \\
		﻿\citeauthor{8735711}~\cite{8735711}	&		&		&		&	$\bullet$	&		&	Hyb. \\
		﻿\citeauthor{8630038}~\cite{8630038}	&		&		&		&	$\bullet$	&		&	GA \\
		
		﻿\citeauthor{MartinKC20}~\cite{MartinKC20}	&		&		&		&		&	$\bullet$	&	GA \\
		\bottomrule
	\end{tabular}
	
	\begin{flushleft}
		\scriptsize$^a$ GA: classic genetic algorithm, eGA: enhanced GA, eNSGA2: enhanced NSGA-II, Hyb.: hybrid GA, Par.: parallel GA 
	\end{flushleft}
\end{table}

Resource allocation is the process of selecting specific devices that execute the applications/store the data \citep{9144705}. A set of constraints limits this selection, and the portion of resources assigned to the agents. In fog systems, this process is even more important because of the limited resources of the devices. This section includes papers in this optimization domain (Table~\ref{tab:surveysegunAsignacion}).

\citeauthor{electronics9030474}~\cite{electronics9030474} studied the task assignment problem for the particular case of opportunistic fog radio access network (OF-RANs) in which user devices can collaboratively create virtual fog access points (v-FAP) within computational capacities. The objective is to reduce the energy consumption, execution time, and maximum use of resources while attempting to achieve a certain balance of the load with this last parameter. They studied the use of the MOEA/D algorithm.

\citeauthor{s19061267}~\cite{s19061267} jointly optimized the allocation of tasks to virtual machines and the placement of the virtual machines into physical resources. The optimization occurs at two different but related levels. For task allocation, they proposed the use of a Hungarian algorithm. For the placement of virtual machines, they used a GA. The GA uses a classic weighted sum for the optimization criteria. The objective was to reduce latency and network usage.

\citeauthor{8325027}~\cite{8325027} used a GA to map virtual devices onto physical devices in a fog infrastructure. The GA minimizes the number of physical devices involved, while also reducing the communication time between these virtual devices. The proposed GA is a hybrid approach in which a classical GA uses an initial population generated using a clustering algorithm (K-means).

\citeauthor{10.1007/978-3-030-33509-0_63}~\cite{10.1007/978-3-030-33509-0_63} presented a preliminary study comparing different solutions for task allocation in fog devices. Among others, they considered a classic GA modified to base the selection function in ordering the solutions rather than using the optimization values.

\citeauthor{8947936}~\cite{8947936} defined an architecture in which devices can voluntarily offer their resources to deploy part of a fog infrastructure. Apart from the architecture, they also proposed a memetic algorithm for the dynamic allocation of fog resources. A memetic algorithm is an extension of GA, and it incorporates a local search process. In this study, the memetic algorithm optimized the placement of virtual containers on fog resources. 

\citeauthor{REDDY2020102428}~\cite{REDDY2020102428} proposed a solution for allocating tasks to virtual machines. The virtual machines were deployed in a fog architecture and serve client task requests. They combined a classic GA with a reinforced learning process that balances energy consumption and latency.

\subsubsection{Service placement}

In fog systems, the placement of services/applications is a special case of the resource allocation problem. In the fog computing paradigm, applications are usually defined as services (unique or interrelated) executed on fog devices. These services can be scaled horizontally by creating different instances and allocating them to devices that best improve a specific optimization criterion \citep{brogi2020place}. The sum of the resources allocated to the services must be smaller than the available resources.

\citeauthor{guerrero2019evaluation}~\cite{guerrero2019evaluation} handled the placement and scaling of applications defined as a composition of interrelated services. The objective was to reduce the latency between services with the same application and resource consumption. They studied and compared three solutions: a classic GA and two multi-objective algorithms, NSGA-II, and MOEA/D. The analysis also included the diversity of the resulting population.

\citeauthor{8929234}~\cite{8929234} considered the case of placing services to virtual machines hosted in the fog nodes of the infrastructure. Considering that service execution time depends on the virtual machine on which it is executed, they optimized the execution time of different services and the energy consumed owing to the execution of the said services. To do this, they proposed to combine the characteristics of a GA and PSO, using GA to first obtain new solutions from the combination of two others and, then, apply the PSO operators to the resulting solution, considering the movement of a particle (solution) toward the local and global optimum.

\citeauthor{8812204}~\cite{8812204} studied the case of a non-deterministic fog architecture. For this, they used a Monte Carlo method combined with a GA. The Monte Carlo method offers an exhaustive exploration of service placement solutions for stochastic architectures, and the GA reduces the number of solutions to be evaluated for each exploration step. This study demonstrated how the combination of both algorithms can considerably reduce the computational cost of the problem, and consequently, the inclusion of GA allows for larger problems without losing diversity in the solutions.

\citeauthor{10.1145/3365871.3365892}~\cite{10.1145/3365871.3365892} handled the application placement problem for interconnected components (services), with the aim to optimize energy consumption (both of the network and the devices), economic cost, and execution time. They implemented the NSGA-II algorithm for the optimization process.

\citeauthor{9110359}~\cite{9110359} considered the allocation of monolithic applications. They proposed using the value of information (VoI) as the optimization criterion. VoI refers to the utility offered by a solution. Using this metric, they proposed the use of a GA with adaptive mutations. \citeauthor{9052663}~\cite{9052663} proposed a similar solution to optimize the location of services in a multi-layer routing environment.

\citeauthor{9012671}~\cite{9012671} studied the problem of placing interrelated services with a classic GA. They considered an infrastructure in which services were encapsulated in virtual containers, and these containers were placed in fog devices. The solution aims to minimize response time by considering execution deadlines and resource usage.

\citeauthor{9238236}~\cite{9238236} addressed the service placement problem in fog environments, considering user mobility. They used a classic GA with a probabilistic fitness function, which improved with users’ mobility information. The optimization objectives were power consumption and quality of service, measured as the percentage of requests that satisfy the execution deadline.

\citeauthor{8676263}~\cite{8676263} also considered user mobility as a critical element for optimizing fog infrastructures. They proposed a double-optimization algorithm. First, the algorithm chooses the services to be executed in fog devices, using the Gini coefficient. Subsequently, it decides specific device to place the service. In the second step, they used a classic GA.

\citeauthor{9363254}~\cite{9363254} studied the service placement problem for the specific case of online gaming applications. In this problem, all users of a game session must use the same service instance. Therefore, by considering sets of users rather than independent users, they proposed minimizing the computational cost and service latency. They used a grouping GA (GGA) combined with a PSO algorithm. GGA selects the service location, and PSO determines the resources allocated to the service.

\citeauthor{app9061160}~\cite{app9061160} implemented a simulator for the placement of multi-component applications. The applications were modeled using DAGs. The simulator allows to choose between several placement policies, including a GA. The parallel design of the simulator leverages the benefits of multi-core architectures.

\citeauthor{NATESHA2021102972}~\cite{NATESHA2021102972} addressed the service placement problem in fog architecture by optimizing service time, cost, and energy. They designed an elitist GA that preserves the set of best solutions between the algorithm generations, avoiding altering them with crossover or mutation operations.

\citeauthor{7776569}~\cite{7776569} proposed a distributed GA to improve the allocation of interrelated services. The GA implements a pool-based GA \citep{GONG2015286}. This design implements a centralized element (a MongoDB database) that stores all the solutions and assigns them to parallel processes that generate new solutions.

As mentioned in the resource scheduling section, it is common to organize fog devices into groups that locally manage the infrastructure. \citeauthor{SkarlatNSBL17}~\cite{SkarlatNSBL17,SkarlatKRB018} defined the concept of fog colony, which is a set of fog devices with a master node that orchestrate and manage the service placement. Therefore, the problem of service location is divided into two decisions: first, finding the most suitable fog device colony and, second, selecting the colony device that will execute the service. They used a classic GA to implement the solution.

\citeauthor{9024660}~\cite{9024660} presented the service placement problem for a fog-based IoT system that maximizes the number of sensors assigned to each service. The coverage radius constrains the problem by limiting the sensor that can be assigned to a service. They used an NSGA-II algorithm that maximized the coverage area and reduced the overlap between coverage areas.

\subsubsection{Service migration}

Service migration is a specific case of service placement that considers the previous location of a service. Decision making not only considers a target placement but also an initial one. Migration is successful when the improvements compensate for the inherent migration costs \citep{tocze2018}. Service migration is more critical in fog computing because of the high dynamicity and mobility of users. For example, some applications require migration, as user positions change with respect to the node that provides coverage \citep{rejiba2019survey}. 

\citeauthor{MartinKC20}~\cite{MartinKC20} studied the migration problem for services encapsulated in virtualized containers. Therefore, the problem focused on the migration of containers between fog nodes. A GA identifies a list of candidate nodes for the migration of each container that needs to migrate. The classic GA with a fitness function that minimizes migration time was used.

\subsubsection{Data allocation}

Some fog domains handle a large number of sensors that collect and generate large-scale data. The need to reduce the data transmitted throughout the network makes it necessary to preprocess the data at levels close to the sensors themselves. Jobs are responsible for preprocessing this data to reduce the size of the transmitted data to the cloud level. The nodes on which these jobs run will have a direct influence on system performance.

\citeauthor{8735711}~\cite{8735711} studied the case of data transmission from fog devices to the Cloud, without data preprocessing. Therefore, the device load is determined by the amount of data to be transmitted. The optimization process consists of assigning tasks to fog nodes of the system such that the total time to transmit data between the nodes and cloud is minimized. In any case, the resulting resource allocation must satisfy two constraints: the limits of tasks allocated to a node and the power consumption. To solve this problem, they proposed to use the Hungarian algorithm to generate the initial population, and then to evolve this population using a GA.

\citeauthor{closer19canali}~\cite{a12100201,closer19canali} studied a case where fog devices filter and partially aggregate data, and finally send it to the cloud to complete this data process. Their proposals were based on assigning data flows generated from the sensors to the data preprocessing jobs. The objective was to reduce the total time required to send and process data between the sensors, fog, and cloud devices. In their initial work, they implemented a classic GA \citep{closer19canali}. Subsequently, \citeauthor{a12100201}~\cite{a12100201} performed a deeper analysis of the GA configuration parameters by studying different crossover, mutation, and selection operators.

Similarly, \citeauthor{8630038}~\cite{8630038} studied the placement of tasks that preprocess spatial data. This problem is unique in that data aggregation has a direct relationship with the data collection area. Therefore, the location of the aggregation tasks also defines the resolution level of the aggregated spatial data. To solve this problem, they proposed the use of a GA that optimizes the resolution obtained from data aggregation.

\subsection{Infrastructure deployment}

\begin{table}[t!]
	\caption{Papers within the scope of fog infrastructure optimization.}
	\label{tab:surveysegunInfraestructura}
	\centering
	\begin{tabular}{p{4cm}|cccc}\toprule
		&	Physical&&&\\ &Location	&	Topology	&	Configuration	&	Algorithm$^a$ \\ \midrule
		﻿\citeauthor{10.1002/spe.2631}~\cite{10.1002/spe.2631}	&	$\bullet$	&		&		&	eGA \\
		﻿\citeauthor{8611104}~\cite{8611104}	&	$\bullet$	&	$\bullet$	&		&	Hyb. \\
		
		\citeauthor{HussainBA20}~\cite{HussainBA20}	&		&	$\bullet$	&		&	NSGA2 \\
		
		﻿\citeauthor{8339513}~\cite{8339513}	&		&	$\bullet$	&		&	Hyb. \\
		
		﻿\citeauthor{9117921}~\cite{9117921}	&		&	$\bullet$	&		&	GA \\
		
		﻿\citeauthor{9220179}~\cite{9220179}	&		&	$\bullet$	&		&	Hyb. \\
		
		﻿\citeauthor{9022931}~\cite{9022931}	&		&	$\bullet$	&		&	NSGA2 \\
		
		\citeauthor{Vorobyev_2019}~\cite{Vorobyev_2019,Vorobyev_2020}&&$\bullet$&&GA\\								
		
		\citeauthor{10.1007/978-3-030-62223-7_20}~\cite{10.1007/978-3-030-62223-7_20}	&		&	$\bullet$	&		&	NSGA2 \\

		\citeauthor{BarikDMSM19}~\cite{BarikDMSM19}	&		&		&	$\bullet$	&	GA \\
		\citeauthor{8422660}~\cite{8422660,8254983}&&&$\bullet$&GA\\
		\bottomrule
	\end{tabular}
	
	\begin{flushleft}
		\scriptsize$^a$ GA: classic genetic algorithm, eGA: enhanced GA, eNSGA2: enhanced NSGA-II, Hyb.: hybrid GA, Par.: parallel GA 
	\end{flushleft}
\end{table}

Some studies were not included in the common domains of resource management. We referred to decision-making problems that address the optimization of network topology, access point locations, network links, or node features (dedicated bandwidth, installed processing capacity, etc.). In this section, we present all the jobs that fall within this group (Table~\ref{tab:surveysegunInfraestructura}).

\citeauthor{8422660}~\cite{8422660} studied the configuration of movable base stations (MBS) in disaster events. To do this, they addressed the determination of processing capacity and transmission bandwidth to avoid data transmission degradation while minimizing the computing and transmission costs. For this, they proposed the use of a classic GA with a weighted objective function. In a previous work \citep{8254983}, the same authors partially worked on the same problem from a similar perspective, in which the objective was to minimize the network communication time.

\citeauthor{BarikDMSM19}~\cite{BarikDMSM19} defined an architecture for the transmission, processing, and analysis of geospatial data based on a fog paradigm. They proposed a GA that optimizes the infrastructure cost by considering the node features (quantity, capacity, and ratio). To do this, they defined a classic GA with a fitness function with weights for the different criteria.

\citeauthor{8339513}~\cite{8339513} presented a study for the deployment of a fog infrastructure in an industrial logistics center. This process determined the physical location of the network components in the factory. The objective was to minimize the cost without loss of performance, latency, and coverage. To achieve this, they use a discrete monkey algorithm (DMA), combined with the operators of a GA to optimize a problem with binary decision variables.

\citeauthor{10.1002/spe.2631}~\cite{10.1002/spe.2631} optimized the initial position of mobile sensors in a fog system. They proposed a classic GA with a fitness function evaluated externally using a trajectory simulator for the aforementioned mobile sensors.

\citeauthor{Vorobyev_2019}~\cite{Vorobyev_2019} and \citeauthor{Vorobyev_2020}~\cite{Vorobyev_2020} proposed to define the infrastructure topology of a fog-cloud architecture by using a classic GA. Each article presents the use of this solution in two different domains.

\citeauthor{8611104}~\cite{8611104} optimized the geographical location of fog nodes and the features of the network links. Initially, they implemented the NSGA-II algorithm. In the preliminary results, they observed the algorithm to be efficient; however, there was an uneven distribution of the solution space. Therefore, they improved the solution by combining a PSO algorithm with NSGA-II. PSO was executed in an initial phase to obtain a suitable distribution of the solutions over the solution space, and subsequently, the NSGA-II improved the population.

\citeauthor{VERBA201948}~\cite{VERBA201948} proposed to consider the interdependence of the applications to improve the location of services in the lower nodes of the fog architecture, thereby optimizing communication time and availability. They handled this problem using a classic GA.

\citeauthor{9022931}~\cite{9022931} proposed a general architecture that organizes IoT-fog-cloud architectures in functional domains. A functional domain is a set of fog nodes independently managed by a coordinator node. The NSGA-II algorithm determines the coordinator using the betweenness centrality.

Similarly, \citeauthor{9220179}~\cite{9220179} also proposed the adaptive formation of federated fog, considered as sets of independent and self-managed nodes. They determined the sets using a hybrid algorithm, which is a classic GA that uses machine learning to calculate fitness.

\citeauthor{10.1007/978-3-030-62223-7_20}~\cite{10.1007/978-3-030-62223-7_20} handled localization and vertical scaling of fog servers in an infrastructure dedicated to neuroimaging diagnosis through a multimodal data fusion process. In this process type, the location of the data fusion points is critical for system performance, and it is important that these coincide with the physical location within the network infrastructure of the servers that perform the process. To do this, they proposed the use of an NSGA-II algorithm and compared its performance with an NPGA-II algorithm (niched Pareto genetic algorithm II). The goal of the multi-objective problem was to minimize the total data fusion time and maximize the success rate.

\citeauthor{HussainBA20}~\cite{HussainBA20} studied how to optimize the physical location, capacity, and the number of fog devices that incorporate computing and storage. The objective was to minimize the response time and energy consumption of these elements. They initially implemented integer linear programming (ILP) to subsequently use an NSGA-II algorithm.

\citeauthor{9117921}~\cite{9117921} improved latency, network traffic, cost, and required resources by optimizing the infrastructure deployment in fog domains. The optimization uses a classical GA with a weighted fitness function.

\section{Design of the genetic optimization}
\label{sec:surveyAG}

We also present the works of this systematic review, grouping them by genetic optimization design. This provides a general overview of GAs that are most common and continue to pose research challenges.

Tables~\ref{tab:surveysegunAG1} and~\ref{tab:surveysegunAG2} show the articles according to the design of the GA:

\begin{itemize}
	\item Standard or classic design. Considering three cases: single-objective GA, multi-criteria NSGA-II, or MOEA/D (labeled as \textit{classic GA}, \textit{NSGA-II}, and \textit{MOEA/D}).
	\item Enhanced or extended design. The base design of a classic GA is extended or improved by including new proposals or modifications. We divided the group into two cases: the classic single-objective GA and the NSGA-II ( labeled as \textit{ enhanced GA} and \textit{ enhanced NSGA-II}).
	\item Hybrid design. The GA is combined with another optimization technique or heuristic (labeled as \textit{Hybrid}).
	\item Parallel implementation. Use of parallelization for the execution of the GA (labeled as \textit{ parallel }).
	
\end{itemize}

The last column of these two tables includes acronyms that refer to the scope of the problem in which the GAs have been applied (labeled as \textit{Scope}). Thus, the information in these tables can be linked to Tables~\ref{tab:surveysegunPlanificacion}, \ref{tab:surveysegunOrquestacion}, \ref{tab:surveysegunAsignacion} and~\ref{tab:surveysegunInfraestructura}.

\begin{table}
	\caption{Paper organization in terms of the algorithm design taxonomy. Subsets of articles including standard or classic designs.}
	\label{tab:surveysegunAG1}
	\centering
	\tabcolsep=0.07cm
		\footnotesize
	\begin{tabular}{p{5cm}|c|c|c|l}\toprule
		
		&	Classic GA	&	NSGA-II	&	MOEA/D	&			Scope$^a$
		\\
		\midrule

		\citeauthor{10.1016/j.future.2019.09.039}~\cite{10.1016/j.future.2019.09.039}		&	$\bullet$	&		&				&	SC	TS\\
		
		﻿\citeauthor{10.3389/fphy.2020.00358}~\cite{10.3389/fphy.2020.00358}		&	$\bullet$	&		&				&	SC	TS\\
		
		﻿\citeauthor{s19061267}~\cite{s19061267} 		&	$\bullet$	&		&				& RA G\\

		\citeauthor{8701449}~\cite{8701449} 		&	$\bullet$	&		&				& SC	DS\\

		﻿\citeauthor{BarikDMSM19}~\cite{BarikDMSM19}		&	$\bullet$	&				&		&	IN	CI\\
		
		﻿\citeauthor{10.1145/3287921.3287984}~\cite{10.1145/3287921.3287984,app9091730} 		&	$\bullet$	&				&		&	SC	TS\\
		
		﻿\citeauthor{9012671}~\cite{9012671}		&	$\bullet$	&		&				& RA SP\\
		
		﻿\citeauthor{closer19canali}~\cite{a12100201,closer19canali} 		&	$\bullet$	&				&		& RA DA\\
		
		﻿\citeauthor{9238236}~\cite{9238236} 		&	$\bullet$	&		&				& RA SP\\
		
		﻿\citeauthor{10.1007/978-3-030-33509-0_63}~\cite{10.1007/978-3-030-33509-0_63} 		&	$\bullet$	&				&		& RA TA\\
		
		﻿\citeauthor{8085127}~\cite{8085127} 		&	$\bullet$	&				&		&	SC	DS\\

		﻿\citeauthor{9117921}~\cite{9117921}		&	$\bullet$	&		&				&	IN	T\\
		
		﻿\citeauthor{MartinKC20}~\cite{MartinKC20}		&	$\bullet$	&				&		& RA M\\
		
		﻿\citeauthor{8422660}~\cite{8422660}\cite{8254983}		&	$\bullet$	&				&		&	IN	CI\\

		\citeauthor{NATESHA2021102972}~\cite{NATESHA2021102972}		&	$\bullet$	&		&				&	RA SP\\

		\citeauthor{9306718}~\cite{9306718} 		&	$\bullet$	&		&		&			SC	DS\\
		
		﻿\citeauthor{9140118}~\cite{9140118}		&	$\bullet$	&				&		&	SC	TS\\
		
		﻿\citeauthor{10.1007/978-981-15-0199-9_24}~\cite{10.1007/978-981-15-0199-9_24}		&	$\bullet$	&				&		&	OR	OR\\
		
		﻿\citeauthor{SkarlatNSBL17}~\cite{SkarlatNSBL17,SkarlatKRB018} 		&	$\bullet$	&				&		& RA SP\\
		
		﻿\citeauthor{VERBA201948}~\cite{VERBA201948}		&	$\bullet$	&				&		& RA SP\\
		
		\citeauthor{Vorobyev_2019}~\cite{Vorobyev_2019}\cite{Vorobyev_2020} &	$\bullet$	&				&		&	IN	T\\		
		
		﻿\citeauthor{8630038}~\cite{8630038}		&	$\bullet$	&				&		& RA DA\\
		
		﻿\citeauthor{8676263}~\cite{8676263}		&	$\bullet$	&				&		& RA SP\\
		
		﻿\citeauthor{7545926}~\cite{7545926},		&	$\bullet$	&		&				&	SC	V\\

		\citeauthor{AbbasiPK20}~\cite{AbbasiPK20}		&		&	$\bullet$	&				&	SC	TS\\
		
		﻿\citeauthor{HussainBA20}~\cite{HussainBA20}		&		&	$\bullet$	&				&	IN	T\\
		
		﻿
		
		﻿\citeauthor{10.1145/3365871.3365892}~\cite{10.1145/3365871.3365892} 		&		&	$\bullet$	&				& RA SP\\
		
		﻿\citeauthor{9045361}~\cite{9045361}		&		&	$\bullet$	&				&	SC	V\\
		
		﻿\citeauthor{9024660}~\cite{9024660} 		&		&	$\bullet$	&				& RA SP\\
		
		﻿\citeauthor{9022931}~\cite{9022931}		&		&	$\bullet$	&				&	IN	T\\
		
		﻿\citeauthor{10.1007/978-3-030-62223-7_20}~\cite{10.1007/978-3-030-62223-7_20}		&		&	$\bullet$	&				&	IN	T\\

		﻿\citeauthor{guerrero2019evaluation}~\cite{guerrero2019evaluation}		&	$\bullet$	&	$\bullet$	&	$\bullet$	&		 RA ES\\

		﻿\citeauthor{electronics9030474}~\cite{electronics9030474}		&		&		&	$\bullet$	&			AR	AT\\
		\bottomrule
	\end{tabular}
	

	\begin{flushleft}
		\scriptsize$^a$ \textbf{RA Resource Allocation }: G General, SP Service Placement, TA Task allocation, DA Data allocation, M
		Migration;	
		\textbf{IN Infrastructure}: C Configuration, T Topology, PL Physical location; 	
		\textbf{SC Scheduling}: WS Workflow scheduling, TS Task Scheduling, DS Data scheduling, V Vehicular;	
		\textbf{OR Orchestration}
	\end{flushleft}

\end{table}

\begin{table}
	\caption{Paper organization in terms of the algorithm design taxonomy. Subset of articles including extensions or improvements in classic designs.}
	\vspace{-2em}
	\label{tab:surveysegunAG2}
	\centering
	\vspace{1em}
	\begin{turn}{90}
		\tabcolsep=0.07cm
		\footnotesize
		\begin{tabular}{p{4.5cm}|l|p{3.2cm}|l|l|l}\toprule
			
			&	Enhanced NSGA-II	&	Enhanced GA	&	Hybrid	&	Parallel	&		Scope$^a$
			\\
			\midrule
			﻿\citeauthor{9233987}~\cite{9233987}				&	Discrete coded	&		&		&		&	SC	TS
			\\
			﻿\citeauthor{DEMAIO2020171}~\cite{DEMAIO2020171}				&	Crossover/mutation	&		&		&		&	SC	WS
			\\
			﻿\citeauthor{10.1007/s11277-017-5200-5}~\cite{10.1007/s11277-017-5200-5}				&	Crowding	&		&		&		&	SC	TS
			\\

			\citeauthor{9300090}~\cite{9300090} 				&		&	Multi-population	&		&		&	OR	OR\\
			
			﻿\citeauthor{9052663}~\cite{9052663}				&		&	Adaptative mutation	&		&		& RA SP\\
			
			﻿\citeauthor{8761932}~\cite{8761932}				&		&	Real coded	&		&		&	SC	V\\
			
			﻿\citeauthor{10.1002/spe.2631}~\cite{10.1002/spe.2631}				&		&	Fitness simulated	&		&		&	IN	PL\\
			
			﻿\citeauthor{8717643}~\cite{8717643}				&		&	Penalty	&		&		&	SC	TS\\
			
			﻿\citeauthor{8359780}~\cite{8359780}				&		&	Adaptative mutation/crossover	&		&		&	SC	TS\\
			
			﻿\citeauthor{s19122783}~\cite{s19122783} 				&		&	Multi-crossover	&		&		&	SC	TS\\

			﻿\citeauthor{9110359}~\cite{9110359}				&		&	Adaptative mutation	&		&		& RA SP\\
			
			﻿\citeauthor{8947936}~\cite{8947936}				&		&	Memetic	&		&		& RA G\\
			
			﻿\citeauthor{8377192}~\cite{8377192}				&		&	Real coded	&		&		& RA TA\\
			
			﻿\citeauthor{Wang2018}~\cite{Wang2018} 				&		&	Local search	&		&		&	SC	DS\\

			\citeauthor{8325027}~\cite{8325027}				&		&		&	K-means $\rightarrow$ GA	&		& RA G\\
			
			﻿\citeauthor{9363254}~\cite{9363254}				&		&		&	GGA and PSO	&		& RA SP\\
			
			﻿\citeauthor{8812204}~\cite{8812204} 				&		&		&	Montecarlo + GA	&		& RA SP\\
			
			﻿\citeauthor{10.23919/FRUCT.2017.8250177}~\cite{10.23919/FRUCT.2017.8250177}\cite{8311595} 				&		&		&	GA, PSO, ACO, AS	&		&	SC	WS\\
			
			﻿\citeauthor{LiZSS20}~\cite{LiZSS20}				&		&		&	GA(BAS)	&		&	SC	V\\
			
			﻿\citeauthor{8735711}~\cite{8735711}				&		&		&	Hungarian $\rightarrow$ GA	&		& RA DA\\
			
			﻿\citeauthor{8339513}~\cite{8339513}				&		&		&	DMA(GA)	&		&	IN	T\\
			
			﻿\citeauthor{REDDY2020102428}~\cite{REDDY2020102428} 				&		&		&	GA + RL	&		&	OR	OR\\
			
			﻿\citeauthor{10.1002/dac.4652}~\cite{10.1002/dac.4652}				&		&		&	GA $\leftrightarrow$ ACO	&		&	SC	TS\\
			
			﻿\citeauthor{9220179}~\cite{9220179} 				&		&		&	GA(ML)	&		&	IN	T\\
			
			﻿\citeauthor{9308549}~\cite{9308549}				&		&		&	GA $\rightarrow$ PSO	&		&	SC	TS\\
			
			﻿\citeauthor{8929234}~\cite{8929234} 				&		&		&	GA(PSO)	&		& RA SP\\
			
			﻿\citeauthor{8611104}~\cite{8611104}				&		&		&	PSO $\rightarrow$ NSGA-II	&		&	IN	T\\

			﻿\citeauthor{7776569}~\cite{7776569}				&		&		&		&	MongoDB	& RA SP
			\\
			
			﻿\citeauthor{app9061160}~\cite{app9061160}				&		&		&		&	Multi-core	& RA SP\\
			﻿\citeauthor{9c434213a47d45b486ebc9c416d132d0}~\cite{9c434213a47d45b486ebc9c416d132d0,7867735} 			&		&		&		&	Spark	&	OR	OR
			\\		
			\bottomrule
		\end{tabular}
	\end{turn}
	
	\begin{flushleft}
		\scriptsize$^a$ \textbf{RA Resource Allocation }: G General, SP Service Placement, TA Task allocation, DA Data allocation, M
		Migration;	
		\textbf{IN Infrastructure}: C Configuration, T Topology, PL Physical location; 	
		\textbf{SC Scheduling}: WS Workflow scheduling, TS Task Scheduling, DS Data scheduling, V Vehicular;	
		\textbf{OR Orchestration}
	\end{flushleft}

\end{table}		

Additionally, we present a summarized description of the most important advantages and limitations of each work in Tables~\ref{tab:summarypapers1},~\ref{tab:summarypapers2}, and~\ref{tab:summarypapers3}.

\begin{table}
	\caption{Summary of benefits and limitations of the surveyed papers (I).}
	\label{tab:summarypapers1}
	\centering
\tabcolsep=0.07cm
\scriptsize
	\begin{tabular}{p{3cm}|p{5cm}|p{5cm}}\toprule
		
		&	Benefits	&	Limitations	
		\\
		\midrule

\citeauthor{8325027}~\cite{8325027}	& Considering virtual devices for offloading. & Only one optimization objective.\\

		\citeauthor{AbbasiPK20}~\cite{AbbasiPK20}				&Multi-objective optimization.& Limited evaluation considering other studies.\\

		\citeauthor{10.1016/j.future.2019.09.039}~\cite{10.1016/j.future.2019.09.039} &	Experiments with large scale scenarios	& Single objective optimization.\\
		
		﻿\citeauthor{10.3389/fphy.2020.00358}~\cite{10.3389/fphy.2020.00358}			&Important improvements with a small number of generations.& Single objective optimization.\\
		
		﻿\citeauthor{s19061267}~\cite{s19061267} 		&Joint task allocation and virtual machine placement optimization.& Experiments simulated with CloudSim instead of iFogSim.\\
		
		\citeauthor{9233987}~\cite{9233987} & Good state of the art analysis. & Lack of real parametrization. \\

\citeauthor{9300090}~\cite{9300090} & Good use of QoS requirements. & Experiments with a small size scenario.\\

\citeauthor{8701449}~\cite{8701449} & A development of  optimal-preserving decomposition into the cascade of resources. & Need further considerations about 5G features.\\		
		
		﻿\citeauthor{BarikDMSM19}~\cite{BarikDMSM19}			&Scalable solution.& Performance study based only in server-side metrics.\\

﻿\citeauthor{9052663}~\cite{9052663} & A SDN model for Fog nodes. & Absence of an overhead study. \\
		
				\citeauthor{9363254}~\cite{9363254} & A comparison with 4 algorithms & A small range of values in the setup of the study.\\

		﻿\citeauthor{10.1145/3287921.3287984}~\cite{10.1145/3287921.3287984,app9091730} 				&Extensive comparison with other optimization techniques.& Definition of the experimentation with a small number of nodes.\\
		
		﻿\citeauthor{9012671}~\cite{9012671}				&Problem model considers the composition of containers.& Small number of nodes in the experiments.\\

﻿\citeauthor{8812204}~\cite{8812204} & A combination of Monte Carlo method and GA. & Small number of scenarios.\\

		﻿\citeauthor{closer19canali}~\cite{a12100201,closer19canali} 				&Extensive evaluation of serveral designs of GAs.& Experiments not including simulation/emulation.\\

﻿\citeauthor{DEMAIO2020171}~\cite{DEMAIO2020171} & Real time requirements and big volume of data. & Lack of comparisons with other proposals.\\

		﻿\citeauthor{9238236}~\cite{9238236} 				& Optimization considering mobility of the nodes.  & Use of a simple genetic algorithm. Experimentation with a reduced number of nodes.\\
		
		﻿\citeauthor{10.1007/978-3-030-33509-0_63}~\cite{10.1007/978-3-030-33509-0_63} 				&Address highly dynamic networks considering node mobility and failure/disconnection.& Difficulty to keep the network status up to date at all times.\\
		
				﻿\citeauthor{guerrero2019evaluation}~\cite{guerrero2019evaluation} 				&Extensive evaluation of serveral designs of GAs.& Experiments not including simulation/emulation.\\

﻿\citeauthor{8761932}~\cite{8761932} & Address real time requirements of the tasks. & Lack of real data.\\

﻿\citeauthor{10.1002/spe.2631}~\cite{10.1002/spe.2631}						& Including mobile sensors & No use of real data.\\

		\bottomrule
	\end{tabular}

\end{table}

\begin{table}
\caption{Summary of benefits and limitations of the surveyed papers (II).}
\label{tab:summarypapers2}
\centering
\tabcolsep=0.07cm
\scriptsize
\begin{tabular}{p{3cm}|p{5cm}|p{5cm}}\toprule
&	Benefits	&	Limitations	\\\midrule

		﻿\citeauthor{HussainBA20}~\cite{HussainBA20}				&Deep study of all the solutions in the Pareto set.& Experiments not including simulation/emulation.\\

﻿\citeauthor{electronics9030474}~\cite{electronics9030474} & Define intermediate access points. & A comparison with other techniques is missing.\\

﻿\citeauthor{10.23919/FRUCT.2017.8250177}~\cite{10.23919/FRUCT.2017.8250177}\cite{8311595} & Including security considerations. & Limited number of features in scenario.\\

\citeauthor{8717643}~\cite{8717643} & Wireless consideration using NOMA protocol. & Lack of other aspects of NOMA.\\

﻿\citeauthor{LiZSS20}~\cite{LiZSS20} & A combination of beetle antennae search algorithm and GA. & Limited evaluation with other algorithms.\\

\citeauthor{8735711}~\cite{8735711} & A distributed model with local information. & No considering heterogeneity of the data. \\ 

﻿\citeauthor{8339513}~\cite{8339513} & A combination of discrete monkey algorithm and GA. & There is no variability in the experimentation.\\

﻿\citeauthor{8359780}~\cite{8359780} & A collaborative scheduling of tasks and fog resources. & A discussion about the overhead of the proposal is missing.\\

		﻿\citeauthor{8085127}~\cite{8085127} 				& Definition of a fog architecture along the resource optimization combined with reliability, fault tolerance, privacy.& The design of the GA is slightly explained.\\

\citeauthor{s19122783}~\cite{s19122783} 						& Use of Interference beaconing protocol (LIBP). & Small size of the scenarios in the experiments.\\

		﻿\citeauthor{9117921}~\cite{9117921}				& Experiments considering high scale scenarios. &Multi-objective optimization implemented with a single objective GA.\\
		
		﻿\citeauthor{MartinKC20}~\cite{MartinKC20}				& Proposal of a conceptual framework along the optimization solution.& No use of predictive mobility models.\\

		﻿\citeauthor{10.1145/3365871.3365892}~\cite{10.1145/3365871.3365892} 				&Simulated and real world evaluation.& No fault tolerance for fog devices is considered.\\
		
		﻿\citeauthor{9045361}~\cite{9045361}				&Multi-objective optimization considering mobility and dynamic resources.& No study of the evolution of the solutions along the generations.\\

﻿\citeauthor{7776569}~\cite{7776569} & Good scalability through distributed computing. & It deals with a few resource features and has centralized storage.\\

		﻿\citeauthor{8422660}~\cite{8422660}\cite{8254983}				& Experiments with realistic parameters for disaster events. & Complex simulations are required for the adaptation of the bandwidth parameters.\\

		﻿\citeauthor{9024660}~\cite{9024660} 				&Centralized and distributed orchestrator is considered.& Proposals compared with a random solution.\\
		
\citeauthor{NATESHA2021102972}~\cite{NATESHA2021102972} 						& Use of the Elitism-based genetic algorithm. & Low variability of values in the experimentation.\\

		\citeauthor{9306718}~\cite{9306718} 				& Extensive experimental comparison with other proposed algorithms. High scale size of the experiments.& Design of the GA slightly detailed.\\
		
		﻿\citeauthor{9140118}~\cite{9140118}				& Experimental results obtained by simulation. & Dependencies between tasks are not considered.\\

\citeauthor{9110359}~\cite{9110359} & Definition of the concept of Volume of Information. & Experiment with static environment.\\

		\bottomrule
	\end{tabular}

\end{table}

\begin{table}
\caption{Summary of benefits and limitations of the surveyed papers (III).}
\label{tab:summarypapers3}
\centering
\tabcolsep=0.07cm
\scriptsize
\begin{tabular}{p{3cm}|p{5cm}|p{5cm}}\toprule
&	Benefits	&	Limitations	\\\midrule

		﻿\citeauthor{10.1007/978-981-15-0199-9_24}~\cite{10.1007/978-981-15-0199-9_24}				& Analysis of the fitness function along the generations of the GA. &Multi-objective optimization implemented with a single objective GA.\\
		
		\citeauthor{REDDY2020102428}~\cite{REDDY2020102428} & Optimize the period of fog nodes’ duty cycle. & Low variability in experimentation.\\

﻿\citeauthor{10.1002/dac.4652}~\cite{10.1002/dac.4652} & A combination of GA and Ant Colony Optimation. & Few number of features in the criteria.\\

\citeauthor{8947936}~\cite{8947936} & Memetic algorithm as search method.  & Few range of values in the features of the experimentation. \\

﻿\citeauthor{app9061160}~\cite{app9061160} & Allow dynamic joint optimization and tracking of the energy and delay performance of fog systems. & Lack of scalability and performance studies.\\

﻿\citeauthor{9220179}~\cite{9220179} & Define a fog federation formation mechanism.  & Lack of prediction tests for the data set.\\

﻿\citeauthor{8377192}~\cite{8377192} & Distributed model to obtain low latencies. & No consideration of network failures. \\

		﻿

		﻿\citeauthor{SkarlatNSBL17}~\cite{SkarlatNSBL17,SkarlatKRB018} 				& Definition of fog colonies to simplify the complexity of the optimization.& Limited experiment comparison with other optimization techniques.\\

﻿\citeauthor{9308549}~\cite{9308549} & Five EA in the experimentation. & Slightly improvements with regard the control algorithms.\\

		﻿\citeauthor{9022931}~\cite{9022931}				& Solution considering user mobility and the topology of the infrastructure.& Service reliability is not considered.\\

﻿\citeauthor{10.1007/s11277-017-5200-5}~\cite{10.1007/s11277-017-5200-5}	& A design of resource scheduling.  & Low number of approaches in the evaluation.\\

		﻿\citeauthor{VERBA201948}~\cite{VERBA201948}				& Use Case scenario and parameters of interest based on Industry 4.0 requirements. Performance and application profiles based on experimental data.& No evaluation of migration of the applications.\\
		
		\citeauthor{Vorobyev_2019}~\cite{Vorobyev_2019}\cite{Vorobyev_2020} 		&Definition of an analytical/mathematical model.& Experiments not compared with other research studies.\\

﻿\citeauthor{Wang2018}~\cite{Wang2018} & Management of a large volume of data. & Absence of analysis of other computer models in the state of the art. \\

		﻿\citeauthor{8676263}~\cite{8676263}				&Jointly optimization of resource usage and off-loading.& Cost of migrations is not considered.\\

		﻿\citeauthor{8630038}~\cite{8630038}				&Two-layer data processing architecture for spatial big data clustering. Evaluation in real data set.& Lack of evaluation of the computation time.\\

	\citeauthor{8929234}~\cite{8929234} & Efficient implementation in C+ of an hybrid  algorithm using GA and PSO. & Lack of completeness of cases and resource features in the experimentation.\\	
		
		﻿\citeauthor{10.1007/978-3-030-62223-7_20}~\cite{10.1007/978-3-030-62223-7_20}				& Application to the field of neuroimaging.& No evaluation of the optimization time.\\

\citeauthor{9c434213a47d45b486ebc9c416d132d0}~\cite{9c434213a47d45b486ebc9c416d132d0,7867735} & Consider parallel computing in the treatment of the population. & More exploration of opportunities for incremental replanning and tuning to improve performance.\\				
		
		﻿\citeauthor{7545926}~\cite{7545926}				& Scalable solution for mobile domains.& Validation performed only against non-offloading.\\

﻿\citeauthor{8611104}~\cite{8611104} & Combine the convergence and searching efficiency of NSGA-II and SMPSO. & The proposal does not improves all the criteria.\\

\bottomrule
\end{tabular}
\end{table}

\subsection{Classic single objective GA}

A simple GA design uses a single objective fitness function that returns a scalar value that facilitates the ordering of the solutions. It also uses some of the standard proposals for a set of operations: crossover, mutation, and selection \citep{lim2017crossover}. A mathematical transformation is required when the optimization involves more than one optimization criterion, for example, the use of a weighted sum.

In terms of the evolution of the generations, some designs consider that one generation completely replaces the entire set of previous solutions. However, some designs incorporate the strategy known as elitism, which refers to preserving the best solutions between generations. Elitism avoids losing good solutions and accelerates the convergence of the algorithm \citep{10.1162/106365600568202}. 
Although elitism is an improvement of the basic design of a GA, it is a standard in most of today implementations. We have included papers with an elitism strategy under the category of classic GA \citep{8701449,NATESHA2021102972}.

Table~\ref{tab:surveysegunAG1} shows that the most common solution adopted for the optimization problems in fog computing is the classical approximation of the GA, considering both single objective function and mathematical transformation. There are 27 papers within this group, which represents approximately 40\% of the total.

\subsection{Standard multi-criteria GA}

One of the most popular solutions is the use of the NSGA-II algorithm \citep{10.1007/3-540-45356-3_83}. NSGA-II is probably the most popular GA for solving multi-criteria optimization problems. In this case, the ordering is based on dominance (Section~\ref{optimizacionrecursos}) rather than the fitness function. Specifically, NSGA-II orders solutions using the dominant fronts. Inside each front set, the crowding distance is used for a partial order, prioritizing the solutions that are further away from others.

Table~\ref{tab:surveysegunAG2} reflects that the use of NSGA-II has also been studied in a significant number of works. NSGA-II was used in nine jobs, which represents approximately 12\% of the jobs. Nominally, a second common multi-criteria algorithm, MOEA/D, is also used in some studies, such as \citeauthor{guerrero2019evaluation}~\cite{guerrero2019evaluation} and \citeauthor{electronics9030474}~\cite{electronics9030474}.

\subsection{Extensions to standard GA}

In addition to the standard implementations of the GAs that we have already mentioned, it is common to find partial modifications of these standard GAs. The analysis of the articles shows extensions in two blocks: those based on the classical single-objective GAs and those that use NSGA-II for multi-criteria problems.

The number of papers proposing the use of a modified NSGA-II is very limited. \citeauthor{DEMAIO2020171}~\cite{DEMAIO2020171} modified the crossover and mutation operations. Specifically, they implemented a version of NSGA-II that uses the simulated binary crossover (SBX) \citep{deb1995simulated} and a polynomial mutation \citep{deb1996combined}. \citeauthor{9233987}~\cite{9233987} studied the use of a discrete representation for the solutions. Finally, \citeauthor{10.1007/s11277-017-5200-5}~\cite{10.1007/s11277-017-5200-5} modified the crowding distance calculation. These three works addressed scheduling optimizations.

By contrast, we found a larger number of classic GAs with improvements. In these cases, the most common improvements are in the basic operations (crossing, mutation, and selection) or in the inclusion of new operations, such as local search operations. Basic improvement refers to the probability of the operations being applied. Rather than having a constant value, these algorithms modified the probability based on the evolution of the solutions used, for example, fitness. This allows for faster convergence \citep{10.1145/2996355}. Some studies in this survey used adaptive probability for the mutation \citep{9052663,9110359} or for both mutation and crossover \citep{8359780}.

\citeauthor{s19122783}~\cite{s19122783} modified the crossover operation and, rather than generating two children, they also created a third child based on the mean of the two parents' values. The objective of this type of modification is to increase the diversity of the solution space \citep{10.1002/int.20348}.

Real encoding is another alternative to the GA chromosome. Typically, GAs encode solutions in a binary form using arrays or matrices. However, this type of encoding has disadvantages, such as Hamming cliffs, which refers to the case in which the Hamming distances are long for those encodings that represent very close solutions. For this reason, real number encoding results in a faster convergence to the optimum; however, crossover and mutation operations also require an adaptation \citep{deb2011multi}. Our analysis included the works of \citeauthor{8377192}~\cite{8377192} and \citeauthor{8761932}~\cite{8761932} that implement real coding, with both cases focusing on task planning.

Regarding the population, another way to improve the GA behavior is to create subsets of the solutions, also called multi-populations \citep{COCHRAN20031087}. This strategy consists of creating a split set of solutions that evolve independently. It is also possible to incorporate a process of exchanging solutions between different subgroups. This type of strategy allows assigning specific regions of the solution space to each subset, thus intensifying the optimization search in those areas. Similarly, they facilitate the possibility of running the search process in parallel. \citeauthor{9300090}~\cite{9300090} proposed the use of this type of approach to address the problem of orchestration and composition of services.

Another common modification of GAs is the inclusion of a local search process in each algorithm iteration. The goal of these local searches is to ensure faster solution convergence and stability. Memetic algorithms are solutions based on GAs that include local search processes to improve the performance of the optimization process \citep{moscato1989evolution}. The work of \citeauthor{8947936}~\cite{8947936} includes this local optimization to improve resource allocation. \citeauthor{Wang2018}~\cite{Wang2018} presented another work that implements a local search to reschedule tasks to finish all of them at the same time.

\citeauthor{8717643}~\cite{8717643} extended the GA design by adding new operators. Specifically, they incorporated a penalty function. This type of operation is incorporated to better manage the restrictions posed by a problem on the obtained solutions \citep{mca10010045}.

Some proposals incorporate alternatives for the fitness function, for example, in scenarios where it cannot be established analytically. \citeauthor{10.1002/spe.2631}~\cite{10.1002/spe.2631} used a simulator to evaluate the fitness of the solutions. The simulator is configured using the solution proposed by the GA, and it calculates the metrics that will be used to evaluate the solution.

\subsection{Hybrid GA}

GAs allow improvements by defining or modifying operations, solution coding, or solution organization, as well as by including other optimization techniques and heuristics. When a GA is combined with other techniques, it is considered a hybrid solution. This combination incorporates the advantages of each heuristic \citep{el2006hybrid}.

In general, strategies that combine different heuristics in an optimization process can be classified using two criteria \citep{Talbi2002,6148272}. First, they are classified by the execution flow of the optimization process. Based on this criterion, the first type is the solution executed in sequential phases, called relays. In relay optimization, the result of the first optimization is the input of the second optimization; thus, the optimization is sequential. The second set corresponds to those executed collaboratively. They are executed simultaneously and independently, however, in a coordinated way, for example, by exchanging solutions throughout the optimization process.

Second, hybrid strategies are classified by a second criterion, the degree of coordination between the heuristics. From this perspective, we find low-level or high-level strategies. In the first case, low-level strategies are based on the integration of a meta-heuristic as an improvement of a part or partial task of another meta-heuristic. Thus, one heuristic is a subordinate component integrated into the master meta-heuristic. By contrast, in high-level strategies, the different meta-heuristics are self-contained and self-sufficient, and the coordination process is based on the exchange of information between them.

Hybrid strategies combine these two classification criteria, providing four possible groups. However, there are very few general examples that combine the use of low-level strategies executed by relays \citep{Talbi2002}, and we did not find any of these cases in this systematic review.

For high-level relay strategies, the optimization process is divided into two self-contained phases. Each phase applies two different algorithms, considering the population resulting from the first one as the initial population of the second algorithm. Several works in the field of fog optimization combine heuristics in a high-level relay manner. For example, combining GA and PSO \citep{9308549}, PSO and NSGA-II \citep{8611104}, the Hungarian algorithm and GA \citep{8735711}, and k-means and GA \citep{8325027}, as the first and the second algorithms, respectively.

For high-level collaborative strategies, optimization is performed simultaneously and in a coordinated manner. The exchange of solutions during the optimization process is one of the most common coordination strategies. For example, \citeauthor{10.1002/dac.4652}~\cite{10.1002/dac.4652} proposed the exchange of solutions between a GA and an ACO. Another coordination strategy is based on executing parallel and independent optimization algorithms and subsequently choosing the most appropriate solutions from each solution. For example, \citeauthor{10.23919/FRUCT.2017.8250177}~\cite{10.23919/FRUCT.2017.8250177} and \citeauthor{8311595}~\cite{8311595} combined four different heuristics (GA, PSO, ACO, and AS) and used a test-and-select strategy to choose the final solution.

Collaborative low-level techniques involve a more complex integration of heuristics or algorithms. In these cases, the entire optimization execution flow is redesigned by integrating and interleaving the tasks of each heuristic to be combined. For example, \citeauthor{LiZSS20}~\cite{LiZSS20} included an additional intermediate task to the classic structure of a GA and the evolution operation of a BAS to improve the solutions obtained from selection, crossing, and mutation. Similarly, \citeauthor{8929234}~\cite{8929234} used a GA to control the evolution of the population, however, integrated the particle motion operations of a PSO to accelerate the evolution of solutions. By contrast, \citeauthor{8339513}~\cite{8339513} proposed a general optimization process controlled by the DMA, which integrates genetic crossing and mutation in the DMA. Finally, \citeauthor{9220179}~\cite{9220179} incorporated a machine learning process to calculate fitness, which improved performance during the optimization process.

A set of works can be defined as hybrid solutions by considering the combination of different algorithms or heuristics to optimize different aspects of a system, however, performed simultaneously and coordinately. For example, \citeauthor{9363254}~\cite{9363254} used a grouping GA (GGA) for the location of services and a PSO that determines the resources assigned to the service in the fog device. This technique was also proposed by \citeauthor{REDDY2020102428}~\cite{REDDY2020102428}, where a GA was used to assign resources to virtual machines to be deployed in the fog architecture, and a reinforced learning process was used to predict the duty cycle of each of these virtual machines. Similarly, \citeauthor{8812204}~\cite{8812204} used the Monte Carlo method to explore the possible conditions of a non-deterministic fog architecture that includes variations in the characteristics of the network infrastructure. For each of the explored cases, the optimization process was performed using a GA.

\subsection{Parallel GA}

The execution time of GAs dramatically increases, for example, with bigger optimization problems, customized operators, higher computational-cost operators, larger datasets, or larger number of non-linear restrictions~\citep{cantu1998survey}. The main advantage of parallel GAs (PGA) is to overcome these limitations in terms of the execution times. GAs require the execution of the same operations (crossover and mutation operators, fitness calculation, constraints validation) on multiple isolated subsets of data (individuals of the solution population) and this allows the parallelization of their execution as islands or neighborhoods~\cite{alba1999survey}. But PGAs are much more than just faster implementations~\cite{10.1145/3400031}. PGAs also allow to implement new optimization models, techniques and operators that can take advantage of parallel and heterogeneous platforms~\cite{alba2005parallel, talbi2013metaheuristics} and they can cooperate with other search procedures more easily than sequential GAs~\cite{alba1999survey}. Finally, PGAs also show better search, even if no parallel hardware is used, and  higher efficiency and efficacy than sequential GAs~\cite{alba1999survey}.

Only three paper, of the analyzed ones, adopted a parallel approach for the execution of the algorithms; however, none of them leveraged the distributed nature of the fog architecture. Particularly, two of them proposed the use of tools of a parallel and distributed nature for executing the GA, MongoDB in the case of \citeauthor{7776569}~\cite{7776569} and Spark in \citeauthor{9c434213a47d45b486ebc9c416d132d0}~\cite{9c434213a47d45b486ebc9c416d132d0} and \citeauthor{7867735}~\cite{7867735}; however, all of them used computing resources external to the fog architecture. Finally, \citeauthor{app9061160}~\cite{app9061160} implemented a simulator that incorporates a GA, and this was programmed to leverage the multi-core characteristics of the processors where it executes.

\section{Swarm algorithms for the optimization of fog infrastructures}

Although this study focuses on the use of GAs in fog optimization, it is also interesting to briefly analyse other references in the related field of swarm computation. In particular, we want to present some other researches proposing the use of Ant Colony Optimization(ACO), Particle Swarm Optimization(PSO), Artificial Bee Colony Optimization(ABC), Firefly Algorithm(FA), Bat Algorithm(BA), and Flower Pollination Algorithm(FPA)~\cite{NAYERI2021103078}. The main differences of swarm algorithms are that:  they rely on the collective behaviour of agents, they select the best solutions among all the available solutions, the solution time increases linearly with the population size, and they have a high tendency to premature convergence~\cite{THAKKAR2020100631}.

It is important to note that the number of studies including swarm-based solutions is much smaller than those based on GAs. This is shown in Table~\ref{swarmsurvey}, where the swarm-based solutions are organized in the terms of the previously listed algorithms. All those works deal with different optimization scopes in the field of fog infrastructures, such as, resource allocation~\cite{9107138, 10.1007/978-3-319-98530-5_19, 10.1007/978-3-030-02613-4_38,10.1080/24751839.2017.1356159,Baburao21}, load balancing~\cite{8716467,8997142, 10.1504/IJWGS.2019.099562}, infrastructure deployment~\cite{linchun2020}, scheduling~\cite{8997142,10.1504/IJWGS.2019.099562,8253470,9544399,10.1080/17517575.2017.1304579,9441304,su10062079,8805176,10.1117/12.2580303,9590674,10.1002/cpe.6163,8746271,10.1007/978-3-030-11641-5_27,10.1002/spe.2867,9463602}, service placement~\cite{ADHIKARI2019100053,9098079,GILL2020760,EYCKERMAN2020100237,9345114,Kishor21,9006805,8790850}, or data placement~\cite{8651773}.

\begin{table}
	\caption{Swarm-based optimization for fog infrastructures.}
	\label{swarmsurvey}
	\centering
	\begin{tabular}{p{5.5cm}|p{7cm}}
		\toprule
		\textbf{Algorithm}
		& \\
		\midrule
		Ant Colony Optimization & \cite{10.1080/24751839.2017.1356159} \cite{8805176}  \cite{9590674} \cite{9098079} \cite{GILL2020760} \cite{EYCKERMAN2020100237} \cite{9345114}  \cite{Kishor21} \cite{9006805}  \\ \midrule
		Particle Swarm Optimization & \cite{Baburao21} \cite{10.1002/cpe.6163} \cite{8746271} \cite{10.1007/978-3-030-11641-5_27}  \cite{10.1002/spe.2867}  \cite{9463602} \cite{8790850} \\ \midrule
		Artificial Bee Colony Optimization &    \cite{8997142} \cite{10.1504/IJWGS.2019.099562} \cite{9544399} \cite{10.1080/17517575.2017.1304579} \cite{9441304} \cite{su10062079} \cite{8651773} \\ \midrule
		Firefly Algorithm & \cite{10.1007/978-3-030-02613-4_38} \cite{ADHIKARI2019100053} \\ \midrule
		Bat Algorithm & \cite{10.1007/978-3-319-98530-5_19} \cite{linchun2020} \cite{8253470} \\ \midrule
		Flower Pollination Algorithm & \cite{9107138} \cite{8716467} \\ 
		
		\bottomrule
	\end{tabular}
\end{table}


\section{Research challenges}
\label{sect:researchchallenges}

The papers analyzed in this systematic review clearly showed that a GA is a suitable solution for resource optimization problems in fog architectures. Most of these solutions are based on standard GA designs, applying a single-objective GA or NSGA-II in the case of multi-criteria optimizations. This provides room for future research directions that should be pursued, and their challenges deserve special attention from the scientific community.

We first present the current challenges in terms of GA design. Subsequently, we analyze the optimization field, which requires additional efforts to solve the remaining challenges.

\subsection{Exploring additional standard genetic proposals}

There is a wide range of standard designs of GA that, to the best of our knowledge, has not been considered for resource optimization in fog. Among these algorithms, we find some as diverse as vector evaluated genetic algorithm (VEGA), Fonseca and Fleming's multi-objective genetic algorithm (MOGA), niched Pareto genetic algorithm (NPGA), non-dominated sorting genetic algorithm III (NSGA-III), Elitist non-dominated sorting genetic algorithm (ENSGA), distance-based Pareto genetic algorithm (DBPGA), and thermodynamic genetic algorithm (TGA) \citep{ghosh2005evolutionary}.

Most of these genetic designs have been studied for the optimization of cloud infrastructures, for example, in the scope of task scheduling using NPGA~\cite{yue2016improved} or MOGA~\cite{ramezani2015evolutionary} and virtual machine placement with NSGA-III~\cite{parvizi2020utilization}. However, they should be revisited and reevaluated for fog infrastructures because there is not any algorithm that fits all the optimization problems, as the ``No Free Lunch'' theorem states~\cite{585893}. Cloud and fog computing have important differences in terms of heterogeneity, limited and geographical distribution of the computational resources, heterogeneity of the interconnection network, and scale level of the infrastructure. Consequently, further researches are required in the evaluation of all those genetic designs in the field of fog computing.

\subsection{Parallel designs}
\label{sect:challengesparallel}

The number of papers that have considered the use of parallel GAs is limited, and they do not consider the use of standard parallel designs. The common types of parallel GAs considers three cases: a master/slave single population scheme (the master performs the population evolution and the slaves calculate the objective functions), a fine-grained single population (with a single population, where each process performs all the genetic operations, requiring high coordination and a solution partitioning scheme), and coarse-grained multi-populations (in which each process works with a population that evolves independently) \citep{cantu1998survey,alba1999survey,10.1145/3400031,knysh2010parallel}.

None of the analyzed works considered the use of fog device resources to execute the optimization process. In this way, it would be possible to utilize the distributed characteristics of a fog system to parallelize the execution of the proposed GAs. \citeauthor{10.1007/978-3-030-00374-6}~\cite{10.1007/978-3-030-00374-6} performed a preliminary study on executing GAs for the optimization of classic problems in edge computing environments (with mobile devices and very heterogeneous characteristics). Their results are promising and, despite the differences with edge environments, their conclusions can be transferred to fog infrastructures, particularly those related to the heterogeneity of the devices. In any case, fog devices would offer more extensive resources than edge devices and would not be as dependent on batteries and energy consumption, which are factors that seem to be determined according to the conclusions of the aforementioned study.

For these reasons, it is worth exploring the possibilities of leveraging the computational resources of fog architectures for executing the optimization process. This would be particularly interesting in the case of parallel GAs from coarse-grained population subgroups.

Finally, the definition of new genetic design should consider the integration and deployment requirements of fog infrastructures. These new solutions need to be easily integrated into distributed and self-organizing orchestrators, brokers, and application managers~\cite{venticinque2019methodology,10.1002/smr.2405}. Additionally, new proposals should consider new fog paradigms, such as osmotic computing~\cite{villari2016osmotic}.

\subsection{Hybrid proposals}

The principle of ``No Free Lunch'' states that no algorithm can fully satisfy all application scenarios~\cite{585893}, requiring the combinations of several alternatives. This review includes several studies that consider hybrid solutions. In any case, because of the multiple design possibilities of this type of algorithm (high or low level, relay, or coordinated) and the wide range of meta-heuristics to combine with \citep{beheshti2013review}, the number of possible cases to study is extremely high, and there is still room for further research.

Optimization processes require both the exploration of the solution space, to guarantee the global optimum, and exploitation, to refine and improve a solution. Population-based heuristics, such as GAs, are effective for exploring; however, they are limited to exploitation. Therefore, it is suitable to combine its use with trajectory-based meta-heuristics, such as hill-climbing, SA, and taboo search. This is precisely one of the points on which future research efforts should focus, particularly in the case of low-level hybrid techniques, which integrate other heuristics in the optimization process of a GA in a subordinate manner.

The combination of population-based meta-heuristics for the design of high-level hybrid solutions, both for relay and coordinated models, is an additional research direction to outperform genetic-based optimizations in fog. For example, PSO requires complex control parameter adjustments that affect the effective particle speed. It can easily fall into local optima, with low convergence accuracy. Consequently, the benefits of hybridization can outperform the performance of GA by combining it with PSO and reducing the execution time and computing cost~\cite{nzanywayingoma2017analysis}.

Population-based optimization algorithms can benefit from ensemble strategies that offer the availability of diverse approaches at different stages and reduce the efforts in the offline setting up of the algorithms~\cite{WU2019695}. In addition, the algorithms can support each other during the optimization process, such that their ensemble results in a versatile and powerful population-based optimization algorithm. The basics of these ensemble processes can perfectly fit the features of fog infrastructure (heterogeneity and geographical distribution). This research line should be investigated by diversifying the set of configuration parameters, clustering structures, search operators, and algorithm designs to create ensemble solutions handling different types of problems. But these future proposals should deal with the limitation of the ensemble strategies~\cite{el2006hybrid}. First, they would require significant efforts in the design of local search because of its high influence on the search performance of the  optimization objective. Second, it is important to balance the exploration and exploitation (global and local search) to solve global
optimization problems.

In addition, considering the previous section on parallel designs (Section~\ref{sect:challengesparallel}), it is extremely necessary to study the parallel implementations of these hybrid solutions \citep{Talbi2002}. To the best of our knowledge, no previous study has considered a parallel version of a hybrid algorithm. This is probably a future research line that would provide significant improvements, particularly for collaborative hybridization cases.

\subsection{Scheduling}

The field of task scheduling has been extensively studied, and these previous results cover optimization needs. By contrast, studies on workflow scheduling are a few, probably because of the complexity of considering task dependency relationships.

Future studies should focus on reducing latency, managing energy consumption, and resource usage, as these are the most critical elements of this type of architecture \citep{10.1145/3325097}. The proposed solutions should explore a wider range of possibilities for the use of GAs.

\subsection{Orchestration}

Service orchestration requires the discovery, selection, and the composition of services. In terms of orchestration, resource management refers to the mechanisms that will optimize the process of selecting and composing services, considering their scalability and, in particular, the heterogeneity of the nodes where they are executed. This is an important difference with cloud orchestration, where details of the node resources are not considered because of their homogeneity. By contrast, fog orchestration needs to incorporate device details in the decision-making process owing to device heterogeneity.

\subsection{Service placement}

The number of articles related to the fog service placement problem is relatively high compared to other optimization areas that were also considered in this study. In any case, most of the proposed solutions are based on standard-weighted GAs or the use of NSGA-II.

Although the three parallel GA proposals that have been found are focused on this optimization field, this type of GA still requires a more in-depth and extensive study for each of the three types of defined parallel GAs, particularly those that use multiple populations.

Additionally, the problem of service scale can be considered part of the placement problem because scalability can be considered as determining the position of the services and the number of instances. Scalability has been studied in a limited number of studies, and more research effort is required.

\subsection{Service migration}

Although service migration can be considered as a specific case of service location, it differentiates in the fact that the previous position of the service has an important influence on the optimization \citep{rejiba2019survey}. Few studies have handled service migration in fog. This problem scope is particularly important in cases in which services need to be located as close as possible to users, and the service migrations follow the user paths, even anticipating migration operations to the user movements \citep{bao2017follow}.

\subsection{Data management}

Computing and data optimization scopes have differentiating characteristics, implying that not all the solutions proposed in one area are applicable to the other. For this reason, data management requires particular and differentiated studies of computing optimization. It has been observed that this interest in data management is latent in the scientific community; however, the number of works is substantially lower than those that focus on computing services, not even reaching 10\% of the total number of papers. Therefore, it is necessary to focus more on optimizing data management using GAs \citep{fi12110190,FEI2019435}.

\section{Conclusions}
Fog computing was initially defined by Bonomi in 2011. His work revealed the benefits of fog in environments with IoT devices that require managing a large volume of data and with real-time or low latency requirements \citep{bonomi2011connected,bonomi2012fog}. Since then, there has been a growing interest in this new computing paradigm, and in recent years, a large number of research papers related to this new technology have emerged. We have targeted this literature review on the use of GAs for resource optimization in fog architectures. 

The main contributions of this paper are as follows: 70 papers obtained from GS and WoS were analyzed, two taxonomies were defined to organize the papers in terms of the optimization scope and the GA design, and the main gaps in the current literature were analyzed to identify future research opportunities.

Related to the first research question, the optimization scope, the areas with the most research challenges are workflow scheduling, service placement, service orchestration, and service migration. Additionally, the interest in data management in fog architectures has resulted in only a few papers in this field. Thus, there is a critical need for future research on optimizations related to data management. 

In terms of the GA design, the second research question, most papers implemented a classic single-objective GA or NSGA-II. Thus, there is still room for future research that considers other standard GA proposals.

Based on our survey findings, three directions are provided for future research with the aim of improving research into fog computing optimization:

\begin{itemize}
    \item Designing parallel and distributed GA that can be easily deployed in highly distributed and heterogeneous environments, such as fog infrastructures. 
    \item Designing hybrid solutions to outperform the limitations of GA in optimization of fog infrastructures. 
    \item Designing genetic-based optimizations that consider the emerging evolutions of fog computing, such us osmotic computing and service adaptation.
\end{itemize}

These future research lines are complementary to the study of other alternative meta-heuristics, such as swarm optimization or other evolutionary algorithms. The design of these other population-based algorithms could also take advantage of their fog architecture to deploy them in a distributed way, sharing the fog devices resources for service execution and optimization tasks. Even, the importance of these meta-heuristics in the optimization of distributed architectures would justify a deep analysis of the current research proposals and the elaboration of a systematic literature review, but due to the high number of studies in this field, this is out of the scope of this paper and remains as future work.

\section*{Funding}
Funding: Project TIN2017-88547-P supported by  the Spanish Goverment (MCIN/ AEI 
/10.13039/501100011033/ ) and the European Commission (FEDER Una manera de hacer Europa).

\section*{Data Availability}

Data sharing not applicable to this article because no datasets were generated or analysed during the current study.

\section*{Author contributions}
All the authors contributed equally to the work.





\bibliographystyle{elsarticle-num-names} 

\bibliography{bibliography}





\end{document}